\documentclass[aps,superscriptaddress,nofootinbib,11pt]{revtex4}
\usepackage[english]{babel}
\usepackage[utf8]{inputenc}
\usepackage{graphicx}   
\usepackage{slashed}
\usepackage{epstopdf}
\usepackage{verbatim}   
\usepackage{color}      
\usepackage{subfigure}  
\usepackage{multirow}
\usepackage{hyperref}   
\usepackage{float}
\usepackage{epsfig,rotating}
\usepackage{amsmath,amssymb}
\usepackage{dsfont}
\restylefloat{table}
\numberwithin{equation}{section}

\newcommand{\be}{\begin{equation}}
\newcommand{\ee}{\end{equation}}
\newcommand{\bea}{\begin{eqnarray}}
\newcommand{\eea}{\end{eqnarray}}
\newcommand{\tilphi}{\widetilde{\phi}}
\newcommand{\tilchi}{\widetilde{\chi}}
\newcommand{\tilsigma}{\widetilde{\Sigma}}
\newcommand{\tilY}{\widetilde{Y}}

\begin{document}
\title{ Kinetic mixing between  a     Higgs and a nearly degenerate Dark scalar: oscillations and displaced vertices.}

\author{Daniel Boyanovsky}
\email{boyan@pitt.edu} \affiliation{Department of Physics and
Astronomy, University of Pittsburgh, Pittsburgh, PA 15260}
\author{Junmou Chen}
\email{juc44@pitt.edu}
\affiliation{Department of Physics and
Astronomy, University of Pittsburgh, Pittsburgh, PA 15260}
\date{\today}

\begin{abstract}
Extensions beyond the Standard Model allow for a gauge singlet scalar to be kinetically coupled with the Higgs. We consider   kinetic mixing between a Dark scalar gauge singlet   \emph{nearly degenerate} with the Higgs,  focusing on the \emph{dynamical} aspects of the mixing phenomena. The renormalization program is carried out by obtaining the one-loop effective action which  yields  an effective non-hermitian Hamiltonian to study the dynamics of mixing. The scalar Higgs becomes a coherent superposition of the mass eigenstates, thus kinetic mixing leads to oscillations and common decay channels in striking similarity with neutral meson mixing. Near degeneracy yields an \emph{enhancement} of the kinetic coupling. For small kinetic mixing we find that the mass eigenstates feature   different lifetimes which result in a wide separation of time scales of evolution along with important coherence   aspects from Dark scalar-Higgs  interference.    The wide separation of scales is manifest as displaced decay vertices which could potentially be a telltale experimental signal of kinetic mixing.

\end{abstract}


\maketitle

\section{Introduction, Motivation and goals}

Although the Standard Model is successful as a description of particle physics on solid experimental grounds, it is clear that an explanation for Dark Matter must be sought in extensions beyond the Standard Model (SM). There are many different alternative proposals for such extensions postulating the existence of one or more new particles that could provide a suitable explanation for Dark matter. One such extension  posits the existence of a \emph{dark sector}, namely one or more new particles  which are singlets under $SU(3)_c\times SU(2)\times U_Y(1)$ and do not couple \emph{directly} to the gauge bosons of the standard model. A sterile neutrino describes a simple dark sector, but there are many other possible alternatives (for recent reviews see\cite{alex,essig}). While preliminary searches for dark sector signals  at Babar\cite{babar} and BESIII\cite{bes} did not report evidence,   search programs to probe dark sectors  at the LHC  are ongoing and various recent studies propose new search directions\cite{alex,essig,cohen} along with complementary searches at  other experimental facilities in various regions of parameters\cite{babar2,belle2,kloe}.

Dark sector particles are envisaged to couple to Standard Model degrees of freedom via a \emph{portal}\cite{alex,essig} the nature of which depends on the spin of the Dark particle: a vector portal is associated with dark photons, a see-saw type mechanism for neutrino mass generation is associated with sterile neutrinos, a simple Higgs portal posits a cubic and cuartic coupling of a Dark scalar to the $SU(2)\times U_Y(1)$ gauge singlet $\Phi^\dagger \Phi$ where $\Phi$ is the Higgs doublet. Direct detection searches at LUX\cite{lux} and XENON\cite{xenon} put severe constraints on the simplest Higgs portals.

Motivated by the potential as a  Dark matter candidate  and the current and forthcoming experimental efforts to elucidate new physics in the Dark sector, we focus here on a different \emph{portal} for a Dark scalar, which in an effective low energy description, gives rise to a \emph{kinetic} mixing between the gauge singlet Dark scalar and the Higgs degree of freedom of the Standard Model. Such a coupling emerges naturally in Randall-Sundrum models of extra dimensions\cite{RS}. Breakdown of scale invariance in these models by the presence of  two branes leads to a dilaton-like degree of freedom, the radion,  which acquires a mass through stabilization\cite{gold}.   The coupling of the Higgs to gravity yields a kinetic mixing between the radion (an $SU(2)\times U_Y(1)$ gauge singlet) and the Higgs\cite{wells,cox,radion5,moshe,csaki,gunion,gunion2} therefore the radion is a Dark scalar candidate via a Higgs portal with kinetic mixing.   The radion mass is not constrained by the scale of compactification, instead it depends on the stabilization scale\cite{gold}, therefore it may be considered a free parameter. This dilaton-like scalar field might well be the lightest particle emerging from higher dimensional extensions beyond the Standard Model. There is a rich phenomenology of radion-Higgs mixing with ongoing searches at the LHC\cite{cox,radion5,moshe,gunion,gunion2,chak,angel,jung,desai,barger,sandes,rizzo}. Here we focus on very different aspects of kinetic mixing between a Dark scalar and the Higgs, namely \emph{dynamical} aspects in the case of \emph{very small kinetic mixing of   nearly degenerate} radion (Dark scalar) and Higgs fields. To the best of our knowledge \emph{dynamical} aspects of kinetic mixing have not yet been explored but  are complementary to the   phenomenology of radion-Higgs mixing motivated by current and future searches at the LHC and linear colliders. The analysis of ref.\cite{sandes} suggests that the LHC constraints on the radion mass $M_\chi$  are weaker for $M_\chi \lesssim 134 \,\mathrm{GeV}$, suggesting that perhaps the radion and  the Higgs are very nearly degenerate in mass. In ref.\cite{desai} an analysis of LHC data rules out a large region of kinetic couplings and radion masses,    leaving, however,  a small  region   for the nearly degenerate case with very small kinetic mixing coupling that is not excluded and where the LHC constraints are weaker. This is precisely the region of radion mass and kinetic coupling that we focus on in this study. The more recent analysis in ref.\cite{gunion} studies LHC constraints for the radion mass in the region $300\,\mathrm{GeV} \leq M_\chi \leq 1\,\mathrm{TeV}$, far larger than the scale that we consider in this article.

\vspace{2mm}

\textbf{Goals:}

Along with the motivation from Dark matter and  searches for new physics at the LHC and future colliders, there is an inherent fundamental motivation to study the \emph{dynamics} in the case when mixing arises via a kinetic term. The kinetic coupling  offers a novel manifestation of mixing, different from the usual momentum independent case such as in neutrino mixing, that \emph{could} yield fundamentally new insights into dynamical aspects of mixing phenomena such as oscillations and coherence. In this article we focus on the case in which the radion and Higgs fields are \emph{nearly degenerate} with very small kinetic mixing coupling a possibility that is motivated by the analysis and results of refs.\cite{chak,sandes,desai} suggesting weaker constraints from LHC data in this region of parameters.

Our main goal  is to describe the \emph{dynamical} aspects of   nearly degenerate Dark scalar  (the radion) kinetically mixed with the Higgs establishing an analogy with both neutrino  and neutral meson mixing. In particular we seek to understand the following aspects: a) the nature of the mass eigenstates, their masses and lifetimes, b) renormalization aspects: kinetic mixing requires a novel renormalization program,  c) aspects of coherence manifest in oscillatory behavior of probabilities as a consequence of interference effects, these are more relevant in the nearly degenerate case, d) space-time propagation of the corresponding mass eigenstates and, in  particular, their decay dynamics and channels studying possible telltale signals such as displaced decay vertices,  seeking to establish potential experimental signals of the Dark sector.

\vspace{2mm}

\section{Models, Field Redefinition and   Mass Matrices}\label{sec:diag}

Kinetic mixing between a  gauge singlet scalar and the Higgs field emerges in the low energy limit of Randall-Sundrum\cite{RS} inspired higher dimensional models, where the dilaton field acquires a mass through stabilization\cite{gold}. In these models the radion (proportional to the dilaton) is the scalar singlet kinetically coupled to the Higgs. After spontaneous symmetry breaking the effective radion-Higgs coupling is given at the bilinear level  by
\be \mathcal{L}_{km} = -\varepsilon \partial_{\mu} \phi \partial^{\mu} \chi \,,\label{rah}\ee we refer to $\chi$ as the Dark-scalar (radion) and $\phi$ as the Higgs fields. As discussed above these models yield a rich phenomenology\cite{wells}-\cite{chak}. We note, without justification of its origin,    that a similar tree-level coupling is obtained by coupling an $SU(3)_c\times SU(2)\times U_Y(1)$ scalar singlet $\chi$ to the standard model Higgs doublet $\Phi$ via a dimension five operator,
\be \mathcal{L}_{km} = -\frac{1}{\Lambda} \,\Big(\partial_{\mu} \chi \Big) \, \Phi^{\dagger} \,\Big(D^{\mu} \Phi\Big) + \mathrm{h.c.} \,,\label{kinmix}\ee where $D^\mu$ is the $ SU(2)\times U_Y(1)$ covariant derivative and $\Lambda$ a high energy scale much larger than the electroweak scale. Upon symmetry breaking and in the unitary gauge it follows that
\be \mathcal{L}_{km} = -\varepsilon   \partial^{\mu} \chi\,\partial_\mu \phi  -\frac{\phi}{\Lambda}\,\partial^{\mu} \chi\,\partial_\mu \phi    ~~;~~ \varepsilon = \frac{v}{ \Lambda} \ll 1 \label{kinmixssb}\ee where $\phi$ is the standard model Higgs field.  The last term featuring a cubic coupling between the Higgs and the radion field may lead to new interactions suppressed by the ratio of the typical energy  scale of the process  to the high energy scale $\Lambda$. We are not aware of current bounds on or phenomenological studies of such coupling. A study of possible bounds on this coupling  from LHC data is beyond the scope of this article.

Within the radion model\footnote{We are considering the case $\varepsilon \ll 1$ therefore neglected a correction to the kinetic mixing from radion field redefinition.}

\be \varepsilon \simeq 6 \,\xi \, \gamma   ~~ \, ,  ~~  \gamma = v/\Lambda_\chi \,,  \label{radioneps} \ee where $v$ is the Higgs vacuum expectation value,  $\Lambda_\chi$ is the vacuum expectation value of the radion field\cite{wells,moshe,csaki} and $\xi$ the coupling to gravity. Conformal coupling corresponds to $\xi \simeq 1/6$. In this model the radion couples to the standard model (SM) degrees of freedom with the
  interaction Lagrangian between $\phi, \chi$ and fermions and massive vector bosons   given for one fermionic species and one massive vector boson  by\cite{wells,moshe,csaki}
\be \mathcal{L}_I = -\Big[Y \overline{\psi}   \psi -\frac{M^2_V}{v} V^\mu V_\mu \Big]\Big(\phi + \gamma \,\chi \Big) \,,\label{LInt}\ee  where $Y$ is the Yukawa coupling. We consider the case $\xi \simeq 1$ with $\varepsilon \ll 1$ and $v/\Lambda_\chi \ll 1$ therefore $\varepsilon \simeq  \gamma $, since this is the region of parameter space in which the latest constraints\cite{chak}  from LHC data along with earlier constraints\cite{desai,sandes} allow for a radion nearly degenerate with the Higgs  with very small radion-Higgs mixing. Consequently we begin our study by first  neglecting the radion coupling to fermions and vector bosons in the interaction Lagrangian (\ref{LInt}) as it is suppressed by $v/\Lambda_\chi\ll 1$.  In section (\ref{sec:radionsm}) we discuss the radiative corrections arising from the coupling of the radion to the (SM) degrees of freedom and their consequences.

We consider the model defined by the Higgs field $\phi$ Yukawa coupled to one fermionic species and kinetically coupled to a \emph{dark} scalar field $\chi$. The   simpler case of a Yukawa coupling will highlight the main physical aspects relevant for our study and will be generalized later to include the contributions to the Higgs self-energy from Standard Model degrees of freedom (see section  \ref{sec:discussion}).

The   Lagrangian density of this  model  is\footnote{We canonically normalized the kinetic term of $\chi$ by a field redefinition\cite{moshe,wells,gunion}.}

\be
\mathcal{L}= \frac{1}{2}(\partial_{\mu}\phi)^2+\frac{1}{2}(\partial_{\mu}\chi)^2-\varepsilon\partial_{\mu}\phi\partial^{\mu}\chi -\frac{1}{2}M_{\phi}^2\phi^2-\frac{1}{2}M_{\chi}^2\chi^2  + \overline{\psi}(i{\not\!{\partial}} -m_\psi)\psi -Y  \overline{\psi} \phi \psi\,.\label{totalL} \ee Without loss of generality we take $\varepsilon >0$\footnote{The Lagrangian is invariant under $\varepsilon \rightarrow -\varepsilon~;~\chi \rightarrow -\chi$}. We will first study the free case $Y=0$ to discuss the main aspects of the diagonalization of the kinetic mixing. Thus, first consider

\be
\mathcal{L}_0[\phi,\chi]= \frac{1}{2}(\partial_{\mu}\phi)^2+\frac{1}{2}(\partial_{\mu}\chi)^2-\varepsilon\partial_{\mu}\phi\partial^{\mu}\chi -\frac{1}{2}M_{\phi}^2\phi^2-\frac{1}{2}M_{\chi}^2\chi^2\,, \label{freeL}  \ee and the field redefinitions

\be
\phi=\frac{1}{\sqrt{2}}(\alpha + \beta) \  \  \  \  \  \ \chi=\frac{1}{\sqrt{2}}(\alpha-\beta)\,. \label{alfabeta}
\ee

The resulting Lagrangian becomes

\bea
\mathcal{L}_0 &= & \frac{1}{2}(\partial_{\mu}\alpha)^2+\frac{1}{2}(\partial_{\mu}\beta)^2-
\frac{\varepsilon}{2}\,[(\partial_{\mu}\alpha)^2-(\partial_{\mu}\beta)^2] \nonumber \\
&-&\frac{1}{2} \Big(\frac{M_{\phi}^2+M_{\chi}^2}{2}\Big) \alpha^2-\frac{1}{2}\Big(\frac{M_{\phi}^2+M_{\chi}^2}{2}\Big)\beta^2- \Big( \frac{M_{\phi}^2-M_{\chi}^2}{2}\Big) \alpha\,\beta \label{def1}
\eea

Now introduce the rescaled fields

\be
\alpha\,\sqrt{1-\varepsilon}=A \ \ \ \ \ \ \  \ \     \beta\,\sqrt{1+\varepsilon} = B \,,\label{rescale}
\ee

leading to

\be
\mathcal{L}_0=  \frac{1}{2}(\partial_{\mu}A)^2+\frac{1}{2}(\partial_{\mu}B)^2-\frac{M_A^2}{2}\,A^2-\frac{M_B^2}{2}\,B^2
-  M_{AB}^2 \,A\,B  \,,\label{ABlag}
\ee where
\be  M^2_A =  \frac{M^2_\phi+M^2_\chi}{2(1-\varepsilon)} ~~;~~M^2_B = \frac{M^2_\phi+M^2_\chi}{2(1+\varepsilon)} ~~;~~ M^2_{AB} =
 \frac{M^2_\phi-M^2_\chi}{2 \sqrt{1-\varepsilon^2} }  \,.\label{ABmasses}\ee
 We note that to avoid tachyonic instabilities and/or fields with negative norms the kinetic mixing parameter is constrained to be $ \varepsilon  < 1$. However, we are interested in the  weak kinetic mixing case with $ \varepsilon  \ll 1$.

After the kinetic mixing is eliminated by the above field redefinition,  we now have   two canonical scalar fields with a non-diagonal  mass matrix. Thus the next step is to diagonalize the mass matrix
\be  \mathds{M} = \left(
                   \begin{array}{cc}
                     M^2_A & M^2_{AB} \\
                     M^2_{AB} & M^2_B \\
                   \end{array}
                 \right)   = \frac{1}{2}\Big(M^2_A+M^2_B\Big)\mathds{I} +
\frac{1}{2}\,\Big[\big(M^2_A-M^2_B \big)^2+ 4 \big(M^2_{AB} \big)^2 \Big]^{\frac{1}{2}}\,
  \left(    \begin{array}{cc}
     \cos(2\theta) & \sin(2\theta) \\
     \sin(2\theta) & -\cos(2\theta) \\
   \end{array}
 \right)  \,,  \label{massmtx}                                                                              \ee where
 \be \cos(2\theta) = \frac{M^2_A-M^2_B}{\Big[\big(M^2_A-M^2_B \big)^2+ 4 \big(M^2_{AB} \big)^2 \Big]^{\frac{1}{2}}}~~;~~\sin(2\theta) = \frac{2\,M^2_{AB}}{\Big[\big(M^2_A-M^2_B \big)^2+ 4 \big(M^2_{AB} \big)^2 \Big]^{\frac{1}{2}}} \,. \label{anglesdef}\ee This mass matrix can be diagonalized by a unitary transformation,
 \be \mathds{M} =  {U}^{-1}(\theta)\,\left(
                                      \begin{array}{cc}
                                        M^2_1 & 0 \\
                                        0 & M^2_2 \\
                                      \end{array}
                                    \right)\, {U}(\theta)\,, \label{rotmmas}\ee where
 \bea M^2_{1} & = & \frac{1}{2} \,\Big\{ M^2_A+M^2_B + \Big[\big(M^2_A-M^2_B \big)^2+ 4 \big(M^2_{AB} \big)^2 \Big]^{\frac{1}{2}}\Big\} \label{mass1} \\ M^2_{2} & = & \frac{1}{2} \,\Big\{ M^2_A+M^2_B - \Big[\big(M^2_A-M^2_B \big)^2+ 4 \big(M^2_{AB} \big)^2 \Big]^{\frac{1}{2}}\Big\}\label{mass2} \eea and the rotation matrix

 \be  {U}(\theta) = \left(
                             \begin{array}{cc}
                               \cos(\theta) & \sin(\theta) \\
                               -\sin(\theta) & \cos(\theta) \\
                             \end{array}
                           \right) \,.\label{rotmtx} \ee In terms of the fields $\phi_1,\phi_2$ that describe the \emph{mass eigenstates}
\be \left(
      \begin{array}{c}
        \phi_1 \\
        \phi_2 \\
      \end{array}
    \right) =  {U}(\theta) \left(
                                    \begin{array}{c}
                                      A \\
                                      B \\
                                    \end{array}
                                  \right) \,,\label{fields12} \ee the   Lagrangian density (\ref{freeL}) becomes
\be \mathcal{L}_0 =  \frac{1}{2}(\partial_{\mu}\phi_1)^2+\frac{1}{2}(\partial_{\mu}\phi_2)^2- \frac{1}{2}M^2_{1}\phi^2_1-\frac{1}{2}M^2_{2}\phi^2_2\,  \label{Lagmass} \ee          and the original fields $\phi,\chi$ are related to the fields that create the mass eigenstates  $\phi_{1,2}$ as

\be\left(
     \begin{array}{c}
       \phi \\
       \chi \\
     \end{array}
   \right) =  \left(
                \begin{array}{cc}
                  y_1 & y_2 \\
                  h_1 & h_2 \\
                \end{array}
              \right)
    ~  \left(
     \begin{array}{c}
       \phi_1 \\
       \phi_2 \\
     \end{array}
   \right)\,, \label{rela}\ee where
   \bea  y_1 & = & \frac{1}{\sqrt{2}} \Big[\frac{\cos(\theta)}{\sqrt{1-\varepsilon}}+\frac{\sin(\theta)}{\sqrt{1+\varepsilon}} \Big]~~;~~ y_2 = \frac{1}{\sqrt{2}} \Big[\frac{\cos(\theta)}{\sqrt{1+\varepsilon}}-\frac{\sin(\theta)}{\sqrt{1-\varepsilon}} \Big] \label{y1y2} \\  h_1 & = & \frac{1}{\sqrt{2}} \Big[\frac{\cos(\theta)}{\sqrt{1-\varepsilon}}-\frac{\sin(\theta)}{\sqrt{1+\varepsilon}} \Big]~~;~~ h_2 = -\frac{1}{\sqrt{2}} \Big[\frac{\cos(\theta)}{\sqrt{1+\varepsilon}}+\frac{\sin(\theta)}{\sqrt{1-\varepsilon}} \Big] \,, \label{h1h2}\eea these coefficients obey
   \be y^2_1 + y^2_2 = h^2_1 + h^2_2 = \frac{1}{1-\varepsilon^2} \,. \label{relayh}\ee This coincides with the field redefinitions and rotations introduced in  refs.\cite{wells,moshe}.

Turning on the Yukawa coupling in (\ref{totalL}) and writing $\phi$ in terms of the mass basis $\phi_1,\phi_2$ is clear that the Yukawa interaction  in  the basis of ``mass eigenstates'' becomes $ \mathcal{L}_Y = -Y\overline{\psi}(y_1\phi_1+y_2\phi_2)\psi $ and both mass eigenstate fields feature Yukawa vertices.   This implies that at one loop there will be $\phi_1-\phi_2$ mixing arising from a self-energy diagram, namely the mass eigenstates couple to a common intermediate state channel. This situation is similar to neutral meson mixing such as $K_0-\overline{K}_0$ mixing, where weak interactions lead to a common intermediate state. Although we focus simply on a Yukawa interaction, this conclusion holds for all the couplings of the Higgs field to the other degrees of freedom of the Standard Model.

The one-loop self energy from the fermion-anti fermion intermediate state features ultraviolet divergences that yield important mass and wavefunction renormalization effects. Rather than studying renormalization in the mass basis, it proves more illuminating to obtain the effective action by integrating out the Fermion fields.

\section{  Effective action and renormalization}\label{sec:effac}
The effective action for the scalar fields can be systematically obtained by carrying out the path integral over the Fermi field. In this path integral the scalar $\phi$ is a passive field acting just like an external field. Consider the functional
\be Z[\phi] = \int \mathcal{D}\overline{\psi}\mathcal{D}\psi~ e^{i\int d^4 x \mathcal{L}[\overline{\psi},\psi;\phi]} \label{Zfun}\ee  with
\be \mathcal{L}[\overline{\psi},\psi;\phi] =  \overline{\psi}(i{\not\!{\partial}} -m_\psi)\psi -Y  \overline{\psi} \phi \psi \,, \label{Lpsi}\ee then
\be \frac{Z[\phi]}{Z[0]} = e^{i\,\delta S_{eff}[\phi]}\,, \label{effac}\ee where $\delta S_{eff}[\phi]$ is the contribution to the scalar effective action from integrating out the Fermionic degrees of freedom. The total effective action is given by
\be S_{eff}=S_0[\phi,\chi] + \delta S_{eff}[\phi] ~~;~~S_0[\phi,\chi]=\int d^4x \mathcal{L}_0[\phi,\chi]\,\label{Seff} \ee where $\mathcal{L}_0[\phi,\chi]$ is given by (\ref{freeL}).

 Normal ordering the Yukawa interaction so that $\langle \overline{\psi}\psi \rangle_\psi =0$ where $\langle \,(\cdots )\,\rangle_\psi$ is the expectation value in the non-interacting Fermion vacuum we find up to order $Y^2$
\be \frac{Z[\phi]}{Z[0]} = 1-\frac{i}{2} \int d^4x_1 \int d^4 x_2 \,\phi(x_1)\, \Sigma(x_1-x_2) \, \phi(x_2) + \cdots \label{Y2eff}\ee where the one-loop self energy
\be i\Sigma(x_1-x_2) =   Y^2 \,\langle \overline{\psi}(x_1) \psi(x_1) \overline{\psi}(x_2) \psi(x_2) \rangle_\psi = (-1)\,Y^2 S(x_1-x_2)\,S(x_2-x_1) \label{sigma}\ee and $S(x-y)$ is the Fermion propagator in position space. Therefore up to one-loop ($\mathcal{O}(Y^2)$) we find
\be \delta S_{eff}[\phi] = -\frac{1}{2}\int d^4x_1 \int d^4 x_2 \,\phi(x_1)\, \Sigma(x_1-x_2) \, \phi(x_2)  \label{delS1lup} \ee

Obviously the effective action is non-local in position space but becomes local in momentum space, introducing the Fourier transforms
\be \phi(x) = \int \frac{d^4 p}{(2\pi)^4} \,  {\tilphi}(p) \,e^{-ip\cdot x} \, \label{FTphi}\ee and similarly for $\chi(x)$ the total effective action up to one loop is given by
\be S_{eff}[\phi,\chi] = \int \frac{d^4 p}{(2\pi)^4} \Bigg\{ \frac{1}{2} \,\tilphi(-p)\,\big[p^2-M^2_\phi-\tilsigma(p) \big]\tilphi(p)+ \frac{1}{2} \,\tilchi(-p)\,\big[p^2-M^2_\chi \big]\tilchi(p) -\varepsilon \, \tilchi(-p)\,p^2 \,\tilphi(p) \Bigg\}\,, \label{totSeff}\ee where for one flavor, and accounting for $N_c=3$ colors the one loop self-energy is given by
\be \tilsigma(p) = -3\,i\,Y^2 \, \int \frac{d^D q}{(2\pi)^D}\,\mathrm{Tr}\Bigg[ \frac{{\not\!{q}}+m_\psi}{q^2-m^2_\psi + i\,0^+}\,\,\,\frac{{\not\!{q}}+{\not\!{p}}+m_\psi}{(q+p)^2-m^2_\psi + i\,0^+}\, \Bigg]   \,. \label{selfie} \ee We calculate this self-energy in dimensional regularization in $D=4-\epsilon, \epsilon \rightarrow 0$ with the result
\be \tilsigma(p) = \frac{3\,\tilY^2}{4\pi^2} \int_0^1 dx\,\Delta[x,p] ~\Bigg\{ 3\,\Big(\frac{2}{\epsilon}-\gamma + \ln[4\pi] + \frac{1}{3} \Big)- 3\,\ln\Big[\frac{\Delta[x,p]}{\mu^2} \Big] \Bigg\} \label{dimregsig}\ee where
\be \Delta[x,p] = m^2_\psi - p^2\,x\,(1-x)-i\,0^+ ~~;~~ \tilY^2 = Y^2\,\mu^{-\epsilon}\,, \label{deltaxp}\ee
and $\mu$ is a renormalization scale. This self-energy features an imaginary part for $p^2 > 4 m^2_\psi$, namely the fermion-anti-fermion threshold with
\be \mathrm{Im}\tilsigma(p) =   - \frac{3\,\tilY^2}{8\pi}\,p^2~\Bigg[1-\frac{4\,m^2_\psi}{p^2}\Bigg]^{\frac{3}{2}}~
\Theta(p^2-4 \,m^2_\psi) \,. \label{imasigma}\ee The real part of $\tilsigma(p)$ is obtained by replacing $\Delta[x,p] \rightarrow \big|\Delta[x,p]  \big|$ in the argument of the logarithm in (\ref{dimregsig}).

\vspace{2mm}

\textbf{Renormalization:} The real part of the self-energy features ultraviolet divergences (a pole in $\epsilon$) that yield mass and wave function renormalization, the latter from the term proportional to $p^2$ in $\Delta[x,p]$. We renormalize on the mass shell of the Higgs field for $\varepsilon=0$, by writing
\be \mathrm{Re}\Big[\tilsigma(p^2)\Big] = \mathrm{Re}\Big[\tilsigma(M^2_{\phi R})\Big] + (p^2-M^2_{\phi R})\,\mathrm{Re}\Big[\tilsigma'(M^2_{\phi R})\Big] + \mathrm{Re}\Big[\tilsigma_f(p^2)\Big] \label{subs} \ee where $\tilsigma'(M^2_{\phi R}) = \partial \tilsigma(p^2)/\partial p^2\Big|_{p^2=M^2_{\phi R}}$. The twice subtracted real part of the self energy   is finite in the limit $\epsilon \rightarrow 0$ and given by
\be \mathrm{Re}\Big[\tilsigma_f(p^2)\Big] \equiv  \mathrm{Re}\Big[\tilsigma(p^2)\Big]-\mathrm{Re}\Big[\tilsigma(M^2_{\phi R})\Big]-
(p^2-M^2_R)\,\mathrm{Re}\Big[\tilsigma'(M^2_{\phi R})\Big]\,. \label{2subsRe}\ee   For $p^2 \simeq M^2_{\phi R}$ it follows that
\be \mathrm{Re}\Big[\tilsigma_f(p^2)\Big] ~~{}_{\overrightarrow{p^2 \rightarrow M^2_{\phi R} }} ~~\mathcal{O}\Big((p^2-M^2_{\phi R})^2\Big) \,.  \label{masshellsigf}\ee

 The mass renormalization condition is
\be M^2_\phi + \mathrm{Re}\Big[\tilsigma(M^2_{\phi R})\Big] \equiv M^2_{\phi R}\,, \label{massren}\ee and introducing the (on-shell) wave function renormalization
\be Z^{-1}_\phi = 1- \mathrm{Re}\,\Big[\tilsigma'(M^2_{\phi R})\Big]\, , \label{zphi}\ee along with  the renormalized field
\be \phi_R = \frac{\phi}{\sqrt{Z_\phi} } \,, \label{renfi}\ee    renormalized mixing
  parameter
\be \varepsilon_R = \sqrt{Z_\phi}\,\,\varepsilon \,, \label{renmix}\ee
 and  the definition
\be \tilsigma_f(p^2) \equiv \mathrm{Re}\Big[\tilsigma_f(p^2)\Big] + i \,\mathrm{Im}\Big[\tilsigma (p^2)\Big] \,,\label{sigmaftot}\ee
  the effective action (\ref{totSeff}) becomes
\bea S_{eff}[\phi,\chi] &=&  \int \frac{d^4 p}{(2\pi)^4} \Bigg\{ \frac{1}{2} \,\tilphi_R(-p)\,\big[p^2-M^2_{\phi R}-Z_\phi\,\tilsigma_f(p^2) \big]\tilphi_R(p)+ \frac{1}{2} \,\tilchi(-p)\,\big[p^2-M^2_\chi \big]\tilchi(p) \nonumber \\ && -\varepsilon_R \,\, \tilchi(-p)\,p^2 \,\tilphi_R(p) \Bigg\}\,. \label{totSeffren}\eea

The factor $Z_\phi$ multiplying $\tilsigma_f$ is absorbed into the renormalization of the Yukawa coupling
\be \widetilde{Y}_R\,Z_Y = \widetilde{Y}\sqrt{Z_\phi}\,Z_\psi \label{Yren}\ee where $Z_Y \simeq (1+ \mathcal{O}(Y^2)+\cdots) $ corresponds to vertex renormalization and $Z_\psi \simeq (1+\mathcal{O}(Y^2)+\cdots) $ to Fermion wavefunction renormalization. Therefore, up to one loop order $Z_\phi \,\tilsigma_f \rightarrow \tilsigma_f$ where in $\tilsigma_f$ the coupling $\widetilde{Y}^2 \rightarrow \widetilde{Y}^2_R$. Finally we choose the renormalization scale $\mu^2=M^2_{\phi R}$ in (\ref{selfie})  according to  on-shell renormalization.

Thus we recognize that the effective action yields a clear renormalization procedure since only the Higgs field $\phi$ undergoes radiative corrections from Standard Model interactions, whereas the $\chi$ field is ``dark'' or ``sterile'' in the sense that it does not couple to the Standard Model degrees of freedom.

There is a clear advantage of the approach to renormalization via the effective action. For consider the alternative of first carrying out field redefinition and diagonalization at the level of the bare action and writing the Yukawa coupling in terms of the bare mass eigenstates. This results in mixed vertices, leading to \emph{six} renormalization conditions: three mass renormalizations  and three wave function-type renormalization conditions,  arising from the divergence proportional to $p^2$ in the self-energy. Two of the three mass renormalization conditions correspond to a mass renormalization of the bare masses and one corresponds to a mass-mixing term. Two of wave-function renormalization conditions  correspond to the wave function renormalizations of the mass eigenstate fields, and one to a kinetic mixing term between the bare fields.  This term leads to a new field redefinition that accounts for the renormalization of the kinetic mixing.  Obviously this alternative manner is much less clear and cumbersome, but ultimately must yield the same results.

This method of renormalization via the one-loop effective action bypasses the more complicated renormalization prescription with mixing\cite{kniehl,pila,pila1}, with the new complications associated with the kinetic mixing term.

\vspace{2mm}

\textbf{Time evolution: no mixing,  $\varepsilon_R=0$:}

Anticipating the discussion of the time evolution for kinetically mixed fields, we first discuss how to extract the effective Hamiltonian in the simpler case of $\varepsilon_R =0$.

For $\varepsilon_R =0$ (no mixing) the $\chi$ field is free and decouples. The renormalized field $\phi_R$ features a propagator
\be G_\phi(p^2) = \frac{-i}{p^2-M^2_{\phi R}- \,\tilsigma_f(p^2)} \,.\label{fiprop} \ee   As a consequence of the on-shell renormalization and the mass-shell behavior of $\mathrm{Re}\Big[\tilsigma_f(p^2)\Big]$ given by (\ref{masshellsigf}) it follows that near the mass shell
\be  G_\phi(p^2 )~~{}_{\overrightarrow{p^2 \rightarrow M^2_{\phi R} }}~~\frac{-i}{p^2-M^2_{\phi R}- \,i \,\mathrm{Im}\Big[\tilsigma_f( M^2_{\phi R})\Big]}\,. \label{nearMS} \ee For $M^2_{\phi R} \gg m^2_\psi$ eqn. (\ref{imasigma}) yields
\be \mathrm{Im}\Big[\tilsigma_f( M^2_{\phi R})\Big] = -M_{\phi R}\,\Gamma_\phi ~~;~~ \Gamma_\phi = \frac{3\,\widetilde{Y}_R^2}{8\pi} \,M_{\phi R} \,,\label{impartMS} \ee where $\Gamma_\phi$ is the decay width at rest of the Higgs scalar (assuming   $M_{\phi R} \gg m_\psi$) into $\overline{f}f$.  Therefore, in absence of kinetic mixing with the dark sector, the renormalization via the effective action yields a Breit-Wigner propagator for the Higgs field $\phi$ near its mass shell
\be  G_\phi(p^2 )~~{}_{\overrightarrow{p^2 \rightarrow M^2_{\phi R} }}~~\frac{-i}{p^2-M^2_{\phi R}+i M_{\phi R}\Gamma_\phi}\,. \label{BWprop} \ee

The main advantage of this renormalization program at the level of the effective action is that neglecting the imaginary part of the self-energy along with kinetic mixing ($\varepsilon_R =0$), the renormalized field $\phi_R$ creates a single particle state with the physical renormalized mass and unit amplitude (unit  residue at the pole in the propagator).   The amplitude of the single particle state of momentum $k$ at time $t>0$ is given by the inverse Fourier transform in $p_0$ of the propagator $G_\phi(p^2=p^2_0-k^2)$. This is dominated by the complex pole in the lower  half $p_0$  plane, namely
\be \mathcal{G}(t,k) = \int \frac{dp_0}{2\pi} \,\frac{e^{-ip_0 t}}{p^2_0-E^2_k+iM_{\phi R}\Gamma_\phi} \propto \, e^{-iE_k t}\,\, e^{-\frac{\Gamma_k}{2}\, t} ~~;~~ E_k = \sqrt{k^2+M^2_{\phi R}}~~;~~\Gamma_\phi(k) = \frac{M_{\phi R}}{E_k}\,\,\Gamma_\phi \,,\label{ampt}\ee the decay rate $\Gamma_\phi(k)$ includes the time dilation factor $1/\gamma(k) = M_{\phi R}/{E_k}$. Anticipating the treatment of the time evolution in the case of mixing in the next section, we re-derive the above result from the effective action for $\phi$ for $\varepsilon_R =0$,

\be S_{eff}[\phi] = \frac{1}{2} \,\int \frac{d^4 p}{(2\pi)^4}     \,\tilphi_R(-p)\,\big[p^2-M^2_{\phi R}- \,\tilsigma_f(p^2) \big]\tilphi_R(p) \,. \label{Seffnomix}\ee The equation of motion for the Fourier transform $\tilphi_R(p)$ in momentum space is
\be \big[p^2-M^2_{\phi R}- \,\tilsigma_f(p^2) \big]\tilphi_R(p) =0 \,,\label{EOMnomix}\ee since $\tilsigma_f(p^2) \propto \widetilde{Y}_R^2 \ll 1$ in perturbation theory the solution is $p^2 = M^2_{\phi R} + \mathcal{O}(\widetilde{Y}_R^2)$ therefore for single particle states of momentum $k$ we write $p_0 = E_k + \delta \omega_k $ with $\delta \omega_k \propto \widetilde{Y}_R^2$ the equation of motion (\ref{EOMnomix}) becomes to leading order in $\widetilde{Y}_R^2$
\be \delta \omega_k \,\tilphi_R(p) = H \, \tilphi_R(p) ~~;~~ H = \frac{\tilsigma_f(p^2=M^2_{\phi R})}{2 E_k }\,.\label{EOMpert}\ee  In terms of the Fourier transform in time,
$\delta \omega_k \rightarrow i \frac{\partial}{\partial t}$ namely $\delta \omega_k$ is associated with the \emph{slow} time evolution of the amplitude through the perturbative self-energy after taking out the trivial phase $e^{-i E_k t}$. Therefore equation (\ref{EOMpert}) becomes a Schroedinger-like equation for the \emph{slow time evolution} of the single particle amplitude,
\be i \,\frac{\partial \tilphi_R(k,t)}{\partial t} = H \, \tilphi_R(k,t)~~;~~ H= \mathrm{Re}\,\Big[ \tilsigma_f(p^2=M^2_{\phi R}) \Big]+ i \,\mathrm{Im}\,\Big[ \tilsigma(p^2=M^2_{\phi R}) \Big]\,,  \label{Schfi}\ee   $H$ is a non-hermitian effective Hamiltonian. Because we renormalized on-shell it follows from (\ref{sigmaftot},\ref{masshellsigf}, \ref{impartMS}) that the\emph{ slow} time evolution of the amplitude obeys
 \be i \,\frac{\partial \tilphi_R(k,t)}{\partial t} = -\frac{i}{2}\,\frac{M_{\phi R}}{E_k}\,\Gamma_\phi \, \tilphi_R(k,t)\,, \label{sloti}\ee namely
 \be \tilphi_R(k,t) = e^{-\frac{\Gamma_\phi(k)}{2} \,t}\,, \ee  which coincides with (\ref{ampt}) after restoring the fast phase $e^{-i E_k t}$. This formulation of the time evolution is equivalent to the Wigner-Weisskopf method ubiquitous in the treatment of the time evolution of neutral meson mixing\cite{lee}-\cite{boya}.

\section{ Mixing: Effective Hamiltonian}\label{sec:effham}

Our strategy to treat the kinetic mixing now begins by writing the total effective action (\ref{totSeffren}) as

\bea S_{eff}[\phi,\chi] &=&  \int \frac{d^4 p}{(2\pi)^4} \Bigg\{ \frac{1}{2} \,\tilphi_R(-p)\,\big[p^2-M^2_{\phi R}\big]\tilphi_R(p)+ \frac{1}{2} \,\tilchi(-p)\,\big[p^2-M^2_\chi \big]\tilchi(p) -   \varepsilon_R \,\, \tilchi(-p)\,p^2 \,\tilphi_R(p)   \nonumber \\ & - & \frac{1}{2} \,  \tilphi_R(-p)\,\,\tilsigma_f(p^2) \,\,\tilphi_R(p)\Bigg\}\,. \label{totSeffren2}\eea
treating the   term $\tilphi_R(-p) \tilsigma_f(p^2) \tilphi_R(p)$   as a \emph{perturbation} and diagonalizing the \emph{first line} in (\ref{totSeffren}) by following the same procedure described in section (\ref{sec:diag}) for the case where $Y=0$ but now in terms of the \emph{renormalized mass, field and mixing parameter}.  The diagonalization of the first line in (\ref{totSeffren2}) follows the same steps as in section (\ref{sec:diag})  with the final result for this contribution to the effective action given by eqn. (\ref{fields12}) but now $M_1,M_2$ and $\tilphi_1,\tilphi_2$ are all in terms of  $\phi_R, M_{\phi,R},\varepsilon_R$. The ``interaction term'' $\tilphi_R(-p) \tilsigma_f(p^2) \tilphi_R(p)$ is now written in terms of the mass eigenstate fields $\tilphi_1,\tilphi_2$ using the relation (\ref{rela}),   namely
\be \tilphi_R(p) = y_1 \tilphi_1(p) + y_2 \tilphi_2 (p) \label{relanew} \ee where $y_1,y_2$ are given by (\ref{y1y2}) but in terms of the renormalized parameters $M_{\phi R},\varepsilon_R$. The total effective action in momentum space in the renormalized $\phi_1-\phi_2$ basis is given by
\bea S_{eff}[\tilphi_1,\tilphi_2] &  = &  \frac{1}{2} \, \int \frac{d^4 p}{(2\pi)^4} \Bigg\{  \,\tilphi_1(-p) \,\big[p^2-M^2_{1}\big]\tilphi_1(p)+  \,\tilphi_2(-p)\,\big[p^2-M^2_2 \big]\tilphi_2(p)\label{Seffmix}\\ & - & y^2_1 \,\tilphi_1(-p) \tilsigma_f(p^2) \tilphi_1(p)   -   y^2_2 \, \tilphi_2(-p) \tilsigma_f(p^2) \tilphi_2(p)-2 \,y_1\,y_2 \,\tilphi_1(-p) \tilsigma_f(p^2) \tilphi_2(p)\Bigg\} \nonumber  \,. \eea

The effective action (\ref{Seffmix}) is fully renormalized and general up to quadratic order in the fields, with the only condition that $|\varepsilon_R| < 1$ to avoid tachyonic instabilities and negative norm states. In this article we focus on the case when the kinetic mixing is very small,     and the Higgs and the Dark scalar are nearly degenerate,   namely we consider
\be M^2_{\phi R} \simeq M^2_\chi ~~;~~ \varepsilon_R \ll 1 \,. \label{dege} \ee It is convenient to  introduce the following parameters
\be \overline{M^2} = \frac{M^2_{\phi R} + M^2_\chi}{2} ~~;~~ \delta = \frac{M^2_{\phi R} - M^2_\chi}{M^2_{\phi R} + M^2_\chi}~~;~~ \eta = \sqrt{\varepsilon^2_R +\delta^2}~~;~~ \eta \,,\, |\delta| \ll 1  \,,\label{masspara} \ee in terms of which we find to leading order in $\varepsilon_R,\delta$
\bea M^2_1  & = &   \overline{M^2}\big[ 1+ \eta\,\big] ~~;~~
M^2_2    =     \overline{M^2}\big[ 1- \eta\,\big] \label{M12small} \\
M^2_{\phi,R} & = & \overline{M^2}\big[ 1+ \delta \big] ~~;~~M^2_{\chi}   =   \overline{M^2}\big[ 1- \delta\big] \label{MfiM} \\
\cos(2\theta) & = & \frac{\varepsilon_R}{\eta} \;;~~~~~~~~~ \sin(2\theta) =\frac{\delta}{\eta} \,.   \label{anglesmall}\eea  For fixed $\delta$ as $\varepsilon_{R} \rightarrow 0 $ it follows that $\theta \rightarrow  \,\mathrm{sign}\,(\delta)\pi/4 $.

At this point one can proceed to obtain the $2\times 2$ propagator matrix in momentum space, the time evolution is then obtained by diagonalizing this matrix and finding the complex poles of the diagonalized propagator. The time evolution is then obtained by performing the Fourier transform in $p_0$, just as described in the previous section. The complex poles at $E_\pm(k) -i \Gamma_{\pm}(k)/2$ with $E^2_\pm(k) = k^2 + M^2_{\pm}$  yield a sum of exponentials
$e^{-iE_\pm(k)t}\,e^{-\Gamma_\pm(k)t/2}$ with coefficients determined by the residues at the poles. In the nearly degenerate case one can take a common rapid phase by writing $E_\pm(k) = \overline{E}(k) + \Delta E_{\pm}(k)$ in terms of the average mass with $|\Delta E_\pm(k)| \ll \overline{E}(k)$ and $\Gamma_{\pm}(k)$ thus describing the slow time evolution. Instead of this procedure, we obtain directly the time evolution from the equations of motion, following the steps of the previous section.

The equations of motion now become
\bea \big[p^2-M^2_1\big]\,\tilphi_1(p) - y^2_1 \tilsigma_f(p^2) \tilphi_1(p)-y_1\,y_2 \tilsigma_f(p^2) \tilphi_2(p) & = & 0 \label{EOMmix1} \\
\big[p^2-M^2_2\big]\,\tilphi_2(p) - y^2_2 \tilsigma_f(p^2) \tilphi_2(p)-y_1\,y_2 \tilsigma_f(p^2) \tilphi_1(p) & = & 0 \,.\label{EOMmix2} \eea The fast time evolution is associated with the scale $ \overline{M^2}$, therefore it proves convenient to write $M^2_1 = \overline{M^2} + \Delta M^2~;~M^2_2 = \overline{M^2} -\Delta M^2$ with $ \Delta M^2=\overline{M^2} \,\eta$ and write
$p_0 = {\overline{E}}_k + \delta \omega_k ~~;~~ {\overline{E}}_k =\sqrt{ k^2+\overline{M^2}}$ where $\delta \omega_k \propto \Delta M^2,\tilsigma_f(p^2=\overline{M^2})$ describes the slow time evolution. Just as we argued in the case without mixing, taking $\delta \omega_k \rightarrow i \partial /\partial t$  the equations of motion (\ref{EOMmix1},\ref{EOMmix2}) become a Schroedinger-type equation for the \emph{slow} time evolution  of coupled channels, namely
\be i\frac{\partial}{\partial t}\, \left(
                                     \begin{array}{c}
                                       \tilphi_1(k,t) \\
                                       \tilphi_2(k,t) \\
                                     \end{array}
                                   \right) = \mathds{H} \, \left(
                                     \begin{array}{c}
                                       \tilphi_1(k,t) \\
                                       \tilphi_2(k,t) \\
                                     \end{array}
                                   \right) ~~;~~ \mathds{H} = \left(
                                                                \begin{array}{cc}
                                                                  H_{11} & H_{12} \\
                                                                  H_{21} & H_{22} \\
                                                                \end{array}
                                                              \right) \, , \label{Scheqnmix}
                                    \ee where
 \bea H_{11} & = &  \frac{1}{2 \overline{E}_k}\,\Big[ \Delta M^2 + y^2_1 \, \tilsigma_f(\overline{M^2})\Big] \label{H11} \\ H_{22} & = &  \frac{1}{2 \overline{E}_k}\,\Big[ -\Delta M^2 + y^2_2 \,\tilsigma_f(\overline{M^2})\Big] \label{H22} \\
 H_{12} & = & H_{21} = y_1\,y_2\,\frac{\tilsigma_f(\overline{M^2})}{2 \overline{E}_k}\,. \label{H1221} \eea

 This is an equation for the \emph{amplitudes}, akin to the effective evolution with a non-Hermitian effective Hamiltonian for amplitudes obtained in the Wigner-Weisskopf formulation of neutral meson mixing\cite{lee,cppuzzle,segre,cpviolation,sanda,boya}. The equivalence between this formulation in terms of the effective non-hermitian Hamiltonian for the amplitudes (Wigner-Weisskopf) and the time evolution obtained from the diagonalization and Fourier transform of the mixed propagator has been established in refs.\cite{sachs,pila1}.

With the definition of $\tilsigma_f(p)$ given by (\ref{sigmaftot}) and with the on-shell renormalization conditions leading to  (\ref{masshellsigf}) it follows that
\be \mathrm{Re}\Big[\tilsigma_f(\overline{M^2}) \Big] \propto \widetilde{Y}_R^2\,\delta^2 \, \overline{M^2}\,, \label{resigM}\ee where we used (\ref{MfiM}). Since $\Delta M^2 = \overline{M^2}\,\eta$ it follows that  for $|\delta| \ll 1$ the contribution from  $\mathrm{Re}\Big[\tilsigma_f(\overline{M^2}) \Big]$ can be neglected, furthermore for $\overline{M^2}\gg m^2_\psi$ we find from (\ref{imasigma})
\be \mathrm{Im}\Big[\tilsigma_f(\overline{M^2}) \Big] = -\frac{3\,\widetilde{Y}_R^2}{8\pi}\, \overline{M^2}\,. \label{imagM}\ee  Hence, defining
\be g_{1,2} = y_{1,2}\,\Big(\frac{\sqrt{3}\,\widetilde{Y}_R}{\sqrt{8\pi}}\Big) \,, \label{g12} \ee the matrix elements of $\mathds{H}$ simplify to
 \bea H_{11} & = &  \frac{\overline{M^2}}{2 \overline{E}_k}\,\big[ \eta - i\,g^2_1\big]   \label{H11s} \\ H_{22} & = &  \frac{\overline{M^2}}{2 \overline{E}_k}\,\big[ -\eta-i\,g^2_2  \big] \label{H22s} \\
 H_{12} & = & H_{21} =  \,\frac{\overline{M^2}}{2 \overline{E}_k}\big[-i\, g_1\,g_2\,\big] \,. \label{H1221s} \eea where $\eta$ has been defined in (\ref{masspara}). In obtaining (\ref{H11s},\ref{H22s}) we neglected contributions of the form (\ref{resigM}) which are subleading in the near degeneracy limit $|\delta| \ll 1$, however they can be incorporated straightforwardly away from this limit.

The effective Hamiltonian $\mathds{H}$ can be written as
\be \mathds{H}= \frac{1}{2} \Big(H_{11}+H_{22} \Big) \,\mathds{I}+ \frac{1}{2} \Big[\Big(H_{11}-H_{22} \Big)^2+ 4 H^2_{12} \Big]^{\frac{1}{2}}~\left(
                                                                  \begin{array}{cc}
                                                                    \mathds{C} & \mathds{S} \\
                                                                    \mathds{S} & -\mathds{C} \\
                                                                  \end{array}
                                                                \right) \label{HeffCS} \ee where
\be  \mathds{C} = \frac{H_{11}-H_{22}}{\Big[\Big(H_{11}-H_{22} \Big)^2+ 4 H^2_{12} \Big]^{\frac{1}{2}}} ~~;~~ \mathds{S} = \frac{2\,H_{12}}{\Big[\Big(H_{11}-H_{22} \Big)^2+ 4 H^2_{12} \Big]^{\frac{1}{2}}}~~;~~ \mathds{C}^2+\mathds{S}^2=1 \,. \label{CandS} \ee Introducing
\be c =   \sqrt{\frac{1+\mathds{C}}{2}}  ~~;~~ s = \frac{\mathds{S}}{\sqrt{2\,\big(1+\mathds{C}\big)} } ~~;~~ c^2+s^2 = 1 \,, \label{smallcs}\ee the effective Hamiltonian $\mathds{H}$ can be diagonalized,
\be \mathds{H} = \mathcal{U}^{-1}\,\left(
                                     \begin{array}{cc}
                                       \lambda_+ & 0  \\
                                       0 & \lambda_- \\
                                     \end{array}
                                   \right)
 \,\mathcal{U} \label{diagH}\ee where the complex eigenvalues are
 \be \lambda_{\pm} =  \frac{1}{2} \Big(H_{11}+H_{22} \Big)  \pm \frac{1}{2} \Big[\Big(H_{11}-H_{22} \Big)^2+ 4 H^2_{12} \Big]^{\frac{1}{2}} \equiv \Delta E_{\pm} - i \frac{\Gamma_\pm}{2} \,,\label{lambdas}\ee and
 \be \mathcal{U} = \left(
                     \begin{array}{cc}
                       c & s \\
                       -s & c \\
                     \end{array}
                   \right) ~~;~~ \mathcal{U}^{-1} = \left(
                     \begin{array}{cc}
                       c & -s \\
                        s & c \\
                     \end{array}
                   \right) \,,\label{mtxU}\ee   because $\mathds{H}$ is non-hermitian, it follows that $\mathcal{U}^{-1} \neq \mathcal{U}^\dagger$. The solution of the effective Schroedinger equation (\ref{Scheqnmix}) is given by
\be  \left(
                                     \begin{array}{c}
                                       \tilphi_1(k,t) \\
                                       \tilphi_2(k,t) \\
                                     \end{array}
                                   \right) =  \left(
                     \begin{array}{cc}
                       c & -s \\
                        s & c \\
                     \end{array}
                   \right)   \left(
                                     \begin{array}{c}
                                       V_+(k,0)~e^{-i \lambda_+ t} \\
                                       V_-(k,0)~e^{-i \lambda_- t} \\
                                     \end{array}
                                   \right) \,, \label{timeevolphi} \ee   where

\be  \left(
                                     \begin{array}{c}
                                       V_+(k,0)  \\
                                       V_-(k,0)  \\
                                     \end{array}
                                   \right) = \left(
                     \begin{array}{cc}
                       c & s \\
                       -s & c \\
                     \end{array}
                   \right)  \,  \left(
                                     \begin{array}{c}
                                       \tilphi_1(k,0) \\
                                       \tilphi_2(k,0) \\
                                     \end{array}
                                   \right)\,. \label{iniconds} \ee Therefore the \emph{slow} time evolution of the amplitudes is given by

\bea \tilphi_1(k,t) & = &  \tilphi_1(k,0)\Big[ c^2 \,e^{-i\lambda_+ t} + s^2 \, e^{-i\lambda_- t}\Big]+ \tilphi_2(k,0)\,c \, s \, \Big[  e^{-i\lambda_+ t} - \, e^{-i\lambda_- t}\Big] \label{fi1oft}\\
\tilphi_2(k,t) & = &  \tilphi_2(k,0)\Big[ s^2 \,e^{-i\lambda_+ t} + c^2 \, e^{-i\lambda_- t}\Big]+ \tilphi_1(k,0)\,c \, s \, \Big[  e^{-i\lambda_+ t} - \, e^{-i\lambda_- t}\Big] \,.\label{fi2oft} \eea  The full time evolution is obtained by multiplying the above amplitudes by the common overall phase $e^{-i\overline{E}_k t}$.  These expressions are similar to those of two-flavor oscillations in neutrino mixing, with \emph{two} major differences: i) the eigenvalues $\lambda_{\pm}$ are \emph{complex}, indicating decay of the amplitudes and ii) the amplitudes $\tilphi_1(k,0),\tilphi_2(k,0)$ are \emph{not independent}, they are determined from the initial amplitudes for the Higgs and dark scalar.

 From the relation (\ref{rela}) we find the amplitudes as a function of time of the original Higgs and Dark scalar fields, namely
\bea \widetilde{\phi}(k,t) & = & y_1\,\tilphi_1(k,t) + y_2 \,\tilphi_2(k,t) \label{higgsoft} \\
\widetilde{\chi}(k,t) & = & h_1\,\tilphi_1(k,t) + h_2 \, \tilphi_2(k,t) \,.\label{chioft}\eea
Finally, we must obtain the initial amplitudes $\tilphi_{1,2}(k,0)$ from the initial amplitudes of the Higgs $\phi$  and Dark scalar $ \chi$ fields. This is achieved by inverting the transformations (\ref{alfabeta},\ref{rescale},\ref{fields12}). We consider the case in which $\tilphi(k,0) \neq 0 ~;~ \tilchi(k,0) =0$ to describe an experimental setting in which a collision has produced a Higgs particle as an initial state, we find
\bea \tilphi_1(k,0) & = &  \tilphi(k,0) \,  F_1 ~~;~~ F_1= \Bigg[\sqrt{\frac{1-\varepsilon_R}{2}}\,\cos(\theta) +  \sqrt{\frac{1+\varepsilon_R}{2}}\,\sin(\theta) \Bigg] \nonumber \\ \tilphi_2(k,0) & = &  \tilphi(k,0)\,  F_2 ~~;~~ F_2 = \Bigg[\sqrt{\frac{1+\varepsilon_R}{2}}\,\cos(\theta) -  \sqrt{\frac{1-\varepsilon_R}{2}}\,\sin(\theta) \Bigg]\,.
\label{inicondis}\eea

Combining (\ref{fi1oft},\ref{fi2oft}), (\ref{inicondis}) with (\ref{higgsoft},\ref{chioft}), and restoring the fast phase $e^{-i\overline{E}_k t}$ we find \bea  \frac{\tilphi(k,t)}{\tilphi(k,0)} & = & e^{-i\overline{E}_k t}\Big[ A_{\phi}\,e^{-i\lambda_+\,t}+ B_{\phi}\,e^{-i\lambda_- \,t} \Big] \label{fiti}\\\frac{\tilchi(k,t)}{\tilphi(k,0)} & = & e^{-i\overline{E}_k t} \Big[A_{\chi}\,e^{-i\lambda_+\,t}+ B_{\chi}\,e^{-i\lambda_- \,t}\Big]\,,  \label{chiti} \eea where
\bea A_{\phi} & = & y_1\,F_1 c^2 + y_2\,F_2\,s^2 + cs\,(y_1\,F_2+ y_2\,F_1) \label{aphi}\\
 B_{\phi} & = & y_1\,F_1 s^2 + y_2\,F_2\,c^2 - cs\,(y_1\,F_2+ y_2\,F_1) \label{bphi} \\
A_{\chi} & = & h_1\,F_1 c^2 + h_2\,F_2\,s^2 + cs\,(h_1\,F_2+ h_2\,F_1) \label{achi}\\
 B_{\phi} & = & h_1\,F_1 s^2 + h_2\,F_2\,c^2 - cs\,(h_1\,F_2+ h_2\,F_1) \,.\label{bchi} \eea

 One finds the identities
\be  y_1\,F_1+y_2\,F_2 =1 ~~;~~ h_1\,F_1+h_2\,F_2 =0 \,,\label{identyhF}\ee which yield
\be A_\phi + B_\phi = 1 ~~;~~ A_\chi + B_\chi =0 \,. \label{ABids}\ee

 The ``disappearance''  $\mathcal{P}_{\phi \rightarrow \phi}(t)$ and ``appearance'' $\mathcal{P}_{\phi \rightarrow \chi}(t)$ probabilities are given by
 \bea \mathcal{P}_{\phi \rightarrow \phi}(t) & = & \Big|\frac{\tilphi(k,t)}{\tilphi(k,0)}\Big|^2 \label{disprob}\\
 \mathcal{P}_{\phi \rightarrow \chi} (t) & = & \Big|\frac{\tilchi(k,t)}{\tilphi(k,0)}\Big|^2 \,.\label{appprob}\eea

 It is convenient to use $\varepsilon_R$ and $\eta = \sqrt{\varepsilon^2_R+\delta^2}$ as parameters, furthermore taking $\phi$ as the standard model Higgs with $M_{\phi} = 125\, \mathrm{GeV}$ it decays into quark-antiquark pairs via the Yukawa couplings to the lower mass quarks, hence the largest Yukawa coupling consistent with Higgs decay is that of the bottom quark, with $3\,\widetilde{Y}_R^2 /8\pi \simeq 4 \,\times 10^{-5}$. We consider $\varepsilon_R \ll 1$, and   the degenerate $\delta =0$ or  the near degenerate cases $|\delta| \ll 1$, therefore $\eta \ll 1$. Furthermore, for $\delta \neq 0$ we will also consider the case where the dark scalar   $\chi$ is heavier than but nearly degenerate with the Higgs, namely $ \delta < 0 $. In this case we find, in terms of $\varepsilon_R,\eta$
\be \cos(\theta) = \sqrt{\frac{1}{2}\Big(1+ \frac{\varepsilon_R}{\eta}\Big)} ~~;~~ \sin(\theta) = - \sqrt{\frac{1}{2}\Big(1- \frac{\varepsilon_R}{\eta}\Big)}\,,\label{sincosteta}\ee  therefore,    for fixed $\delta< 0 $ it follows  from (\ref{y1y2}) that $y_1,y_2$ are   odd  and even functions of $\varepsilon_R$ respectively, consequently, for $ \delta \neq 0 $ (namely $\eta > \varepsilon_R$)  it follows that $y_1 < y_2$.

If $\phi$ is heavier than $\chi$ then $\sin(\theta)$ changes sign and  the odd/even properties  of $y_1,y_2$ as functions of $\varepsilon_R$ are reversed.

Before studying particular cases, we note that
  $ {\varepsilon_R =0~;~\delta \neq 0}$  corresponds to $\theta = - \pi/4$, yielding
\be y_1 = 0 ~,~ y_2 = 1 ~~;~~ h_1 = 1 ~,~ h_2 =0 ~~;~~ F_1=0 ~,~ F_2 = 1 \,,  \label{epszero}\ee which leads to $ \tilphi_2 = \phi~~;~~ \tilphi_1 = \chi$ and $g_1 =0 ~;~ \beta =0 ~~;~~ c=1 ~,~ s=0$. These values yield $ A_\phi =0 ~,~ B_\phi=1 ~~;~~ A_{\chi}=B_{\chi}=0 $. Clearly this is simply the limit in which the kinetic mixing vanishes, the mass eigenstate $\tilphi_2$ is the Higgs field and $\tilphi_1$ is the uncoupled dark scalar field. The rotation by $\theta = -\pi/4$ obviously un-does the (unitary) transformation (\ref{alfabeta}).

\vspace{2mm}

\section{Time evolution: long and short lived modes, oscillations and displaced vertices}\label{sec:timevol}
 It is straightforward to study the general cases numerically, however, there are two relevant cases that can be studied analytically and offer detailed insights into the dynamics.

 \vspace{2mm}

 \subsection{Case I: $\eta \gg g^2_{1,2}$}\label{subsec:caseI}  more precisely this case corresponds to \be \frac{3\,\widetilde{Y}_R^2}{8\pi} \simeq  4\times 10^{-5}  \ll \eta \ll 1 ~ \Rightarrow ~ g^2_{1,2}/\eta \ll 1 \label{ratiogeta} \ee where $g_{1,2}$ are given by (\ref{g12}). To leading order in $g_{1,2}/\eta$ we find
\bea  \mathds{S}   & \simeq &   -i\,\beta   ~~;~~ \mathds{C}   \simeq    1+ \frac{\beta^2}{2}  ~~ ; ~~ \beta = \frac{g_1 g_2}{\eta} = \frac{3\,\widetilde{Y}_R^2}{8\pi\,\eta}\,y_1\,y_2  \ll 1 \label{capSC}  \\
s & \simeq & -i \,\frac{\beta}{2} ~~;~~ c \simeq  1 + \frac{\beta^2}{8} \, ,\label{smallsc}\eea along with the eigenvalues
\bea \lambda_+  & \simeq & \frac{\overline{M^2}}{2 \overline{E}_k}\,\big[ \eta - i\,g^2_1\big] ~~;~~ g^2_1 = \frac{3\,\widetilde{Y}_R^2}{8\pi}\,y^2_1 \label{lamplu}\\ \lambda_-  & = & \frac{\overline{M^2}}{2 \overline{E}_k}\,\big[ -\eta - i\,g^2_2\big] ~~;~~ g^2_2 = \frac{3\,\widetilde{Y}_R^2}{8\pi}\,y^2_2\,. \label{lammin} \eea With $\Delta E_\pm, \Gamma_\pm$ defined by eqn. (\ref{lambdas}),   these are given, in this case by
\bea \Delta E_+ & = &  \frac{\overline{M^2}}{2 \overline{E}_k}\,  \eta ~;~ \Gamma_+ (k) = \frac{\overline{M^2}}{ \overline{E}_k}\,   \,g^2_1 \label{EpGp} \\
 \Delta E_- & = &  -\frac{\overline{M^2}}{ 2 \overline{E}_k}\,  \eta ~;~ \Gamma_- (k) = \frac{\overline{M^2}}{2 \overline{E}_k}\,   \,g^2_2 \,, \label{EmGm} \eea
 the inequality (\ref{ratiogeta}) physically means that $\Delta E_{\pm}(k) \gg \Gamma_{\pm}(k)$, namely in terms of the poles in the propagator, the difference in the real part of the poles is much larger than the individual (and sum of the) widths. Therefore the complex poles in the propagator are spectrally resolved and describe well separated resonances.

  It is convenient to introduce the dimensionless variable
 \be \tau_k = \frac{\overline{M^2}}{  \overline{E}_k} \,t   = \frac{\sqrt{\overline{M^2}}}{\overline{\gamma}_k} \, t \label{tau} \ee where $\overline{\gamma}_k$ is the average Lorentz time dilation factor,  in terms of which the probabilities are given by
 \bea   \mathcal{P}_{\phi \rightarrow \alpha}(t) &    = &   \big| A_\alpha\big|^2 \,e^{-g^2_1 \tau_k}+ \big|B_\alpha \big|^2 \,e^{-g^2_2 \tau_k}  +   2\,e^{-(g^2_1+g^2_2)\tau_k/2} \times \nonumber \\ & & \,\Bigg\{ \big[ A_{R \alpha} B_{R \alpha} + A_{I \alpha} B_{I \alpha}] \cos(\eta \tau_k) + \big[ A_{I \alpha} B_{R \alpha} - A_{R \alpha} B_{I \alpha}] \sin(\eta \tau_k)\Bigg\}\,, \label{probas}\eea where $R, I$ stand for the real and imaginary parts respectively and   $\alpha = \phi, \chi$ respectively.  We note that $\eta \tau_k = [(M^2_1-M^2_2)/2\overline{E}_k ] \, t$ and  $g^2_{1,2} \, \tau_k \equiv \Gamma_{+,-}(k)\,  t$  where $\Gamma_{\pm}(k)$ given by (\ref{EpGp},\ref{EmGm}) are   the decay rates of the corresponding mass eigenstates including the (average) time dilation factor. The  oscillations in (\ref{probas})  are a result of interference between the mass eigenstates, reflecting the fact that mixing between the Higgs and the Dark scalar entails that the Higgs field is a coherent superposition of mass eigenstates, just like ``flavor'' states in neutrino oscillations.

  In the limit (\ref{ratiogeta}) described by Eqns. (\ref{capSC}-\ref{smallsc}), the terms with $s^2, c^2$ are real and only the terms with the product $cs \simeq -i\beta/2$ are imaginary. This substantially simplifies the expressions for the probabilities.

  Before we engage in a numerical study of time evolution of the probabilities, it is illuminating to understand analytically some limiting cases. We consider the case $M^2_\chi  \geq  M^2_\phi$ corresponding to $\delta \leq  0$.

\vspace{2mm}

\subsubsection{$\varepsilon_R \ll 1$, $\delta =0$. }\label{subsec:case11}

  This case corresponds to the Higgs and Dark scalars being degenerate, we refer to this as the \emph{resonant} case.  With $\eta = \varepsilon_R \ll 1 $  we find to leading order in $\varepsilon_R$  that $ y_1 = y_2 = h_1 = h_2 = F_1 = F_2 =1/ \sqrt{2}$  yielding $g_1 = g_2$. Kinetic mixing splits the degeneracy of the mass eigenstates because $\eta =\varepsilon_R \neq 0$ but in this case with $\eta \gg g^2_{1,2}$  both mass eigenstates  feature the \emph{same decay rate} $\Gamma_+ = \Gamma_- = \Gamma_\phi/2 $ where $\Gamma_\phi = 3 \widetilde{Y}_R^2 \,M_{R \phi}/8\pi$ is the Higgs decay rate into the $\overline{f}f$ channel. While this \emph{resonant} case is interesting because it results in an \emph{enhancement} of the kinetic coupling as the   degenerate mass eigenstates  feature the \emph{same} decay widths, it is experimentally ruled out because in this case the Higgs mode features \emph{half} the lifetime of the Standard Model Higgs.

 \vspace{2mm}

 \subsubsection{$\varepsilon_R \ll 1$, $|\delta| \simeq \varepsilon_R$. }\label{subsec:case12}

\begin{figure}[ht!]
\begin{center}
\includegraphics[height=4in,width=3.2in,keepaspectratio=true]{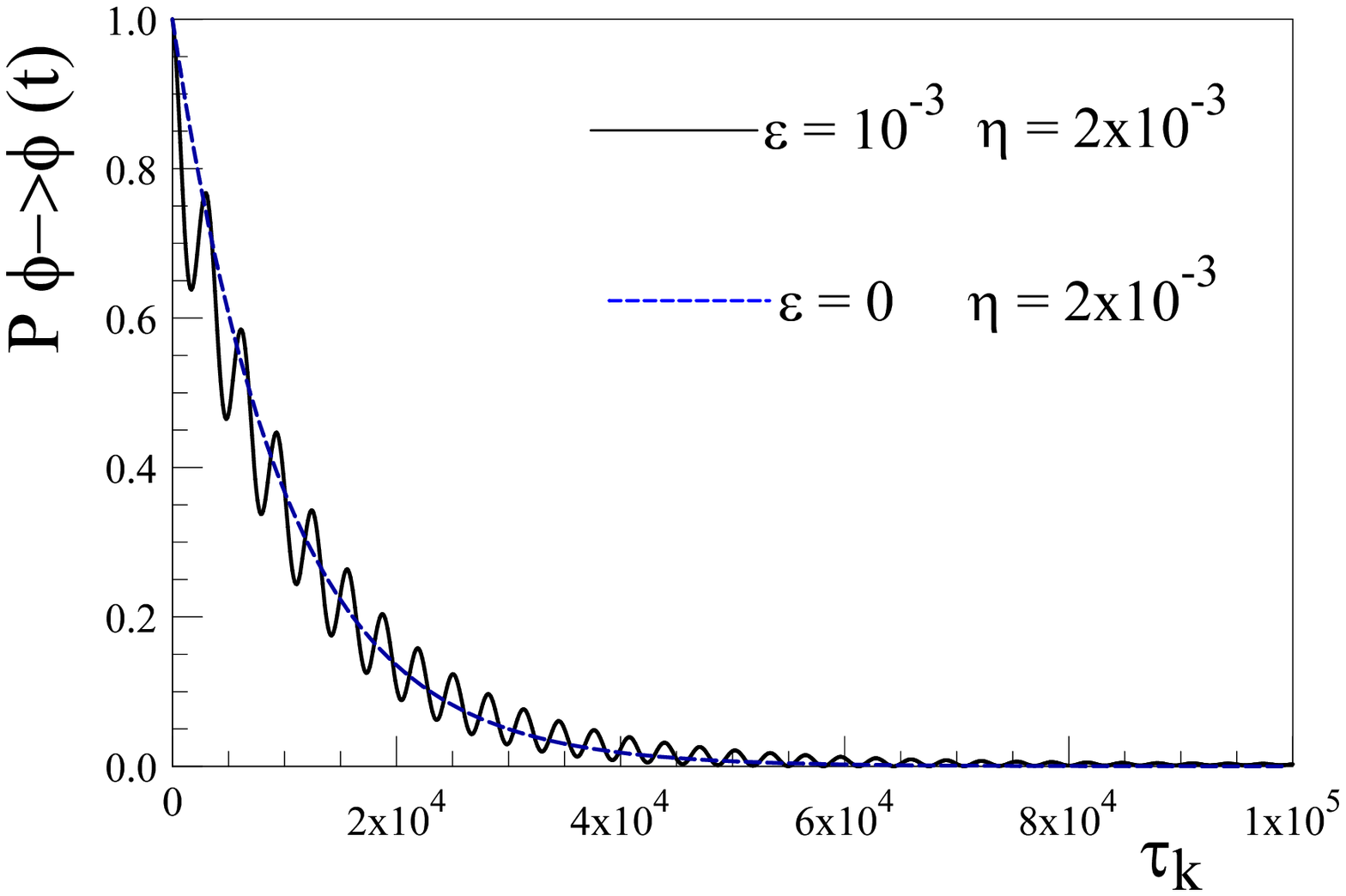}
\includegraphics[height=4in,width=3.2in,keepaspectratio=true]{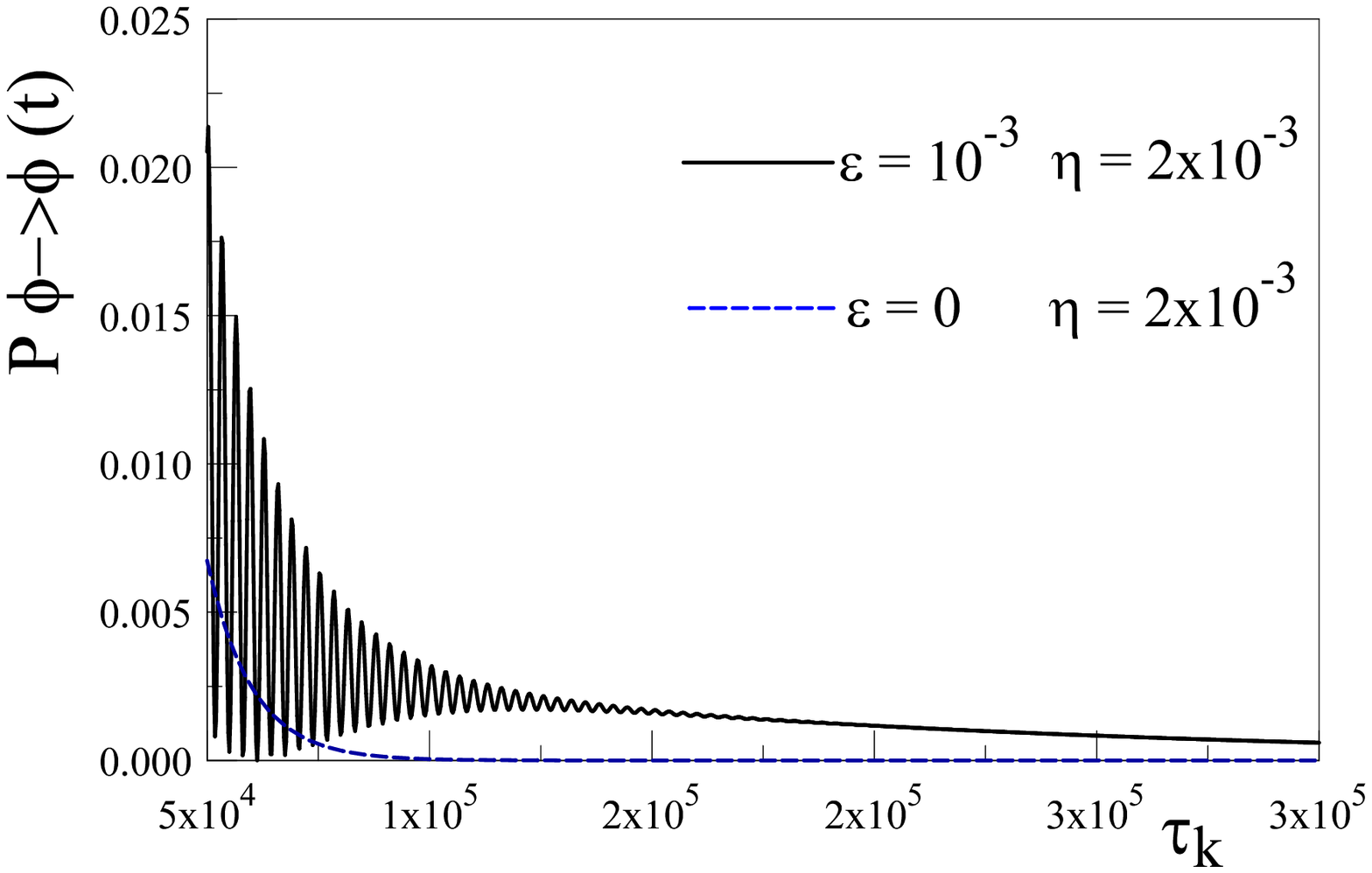}
\includegraphics[height=4in,width=3.5in,keepaspectratio=true]{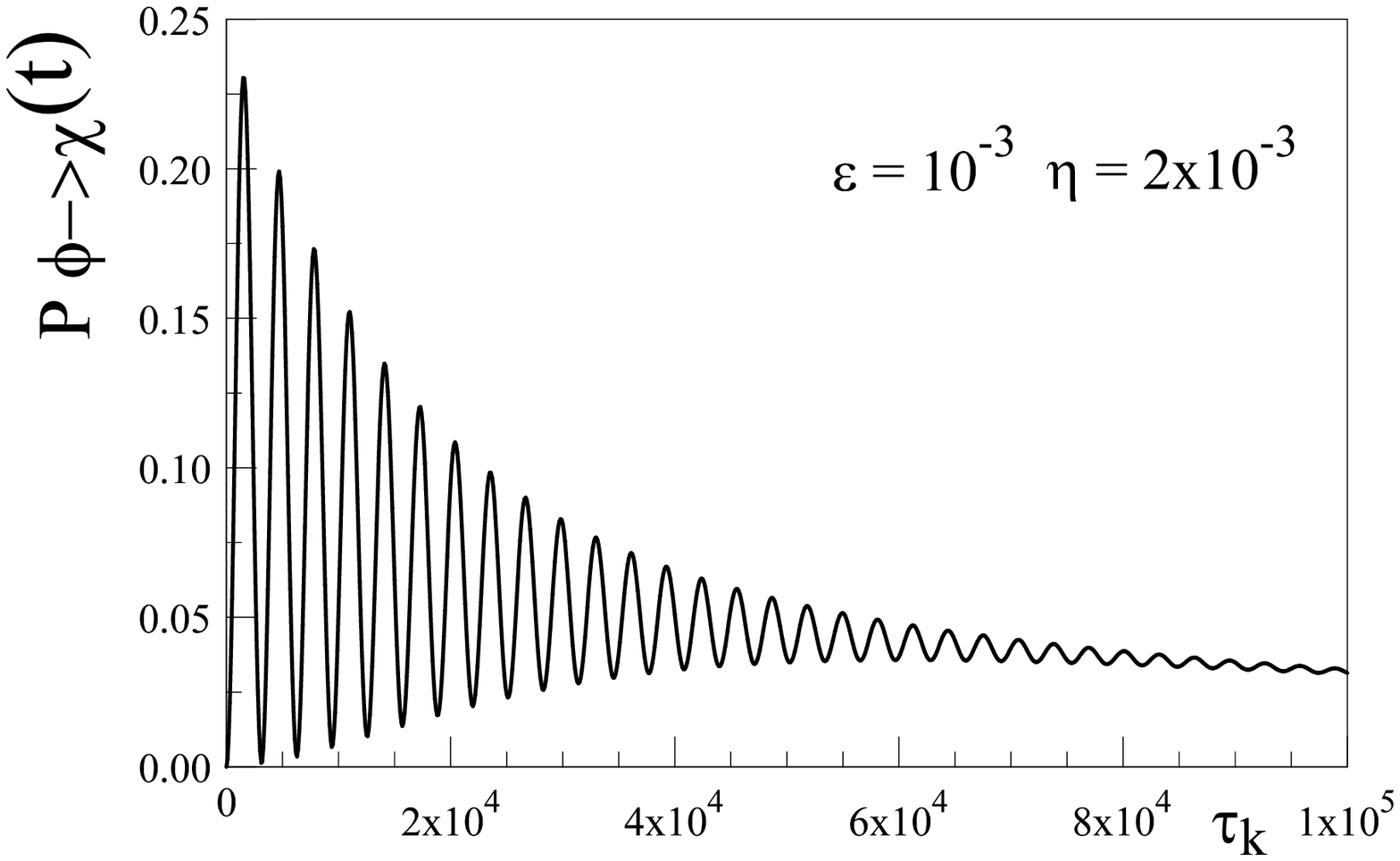}
\caption{ $\mathcal{P}_{\phi \rightarrow \phi}(t) $, the solid line corresponds to $\varepsilon_R = 10^{-3}~,~\eta=2\times 10^{-3}$, the dashed line shows Higss decay without mixing ($\varepsilon_R=0$). The rightmost figure displays the dynamics on the longest time scale. The lowest figure displays $\mathcal{P}_{\phi \rightarrow \chi}(t)$. }
\label{fig:fiproba12}
\end{center}
\end{figure}

\begin{figure}[ht!]
\begin{center}
\includegraphics[height=4in,width=4.5in,keepaspectratio=true]{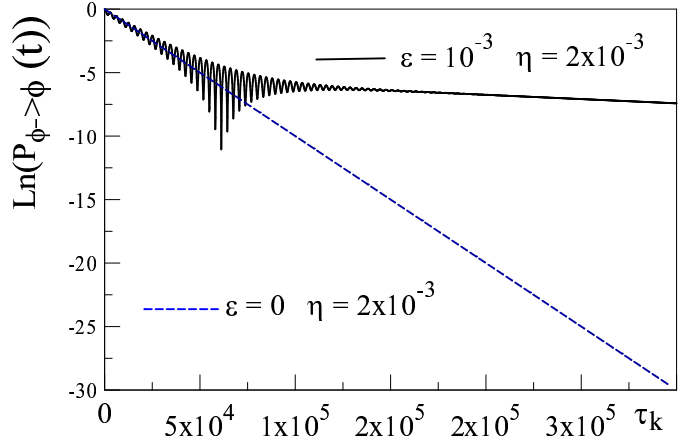}
\caption{ $\ln\Big(\mathcal{P}_{\phi \rightarrow \phi}(t)\Big) $, the solid line corresponds to $\varepsilon_R = 10^{-3}~,~\eta=2\times 10^{-3}$, the dashed line shows Higss decay without mixing ($\varepsilon_R=0$).  }
\label{fig:filogpro}
\end{center}
\end{figure}

 In this case the Higgs and Dark scalars are \emph{nearly} but not degenerate. For $|\delta| \simeq \varepsilon_R$ it follows that $\eta \simeq \varepsilon_R$, this intermediate parameter range must be studied numerically. As an example, the  probabilities $\mathcal{P}_{\phi \rightarrow \phi}$ and $\mathcal{P}_{\phi \rightarrow \chi}$ for $\varepsilon_R = 10^{-3}~,~\eta=2\times 10^{-3}$ are displayed in figs. (\ref{fig:fiproba12},\ref{fig:filogpro}). These parameters describe  the almost degenerate case with $M_\chi/M_\phi = 1.002$. Figs. (\ref{fig:fiproba12},\ref{fig:filogpro}) reveal  the short and long time scales along with the oscillatory behavior, with $g^2_1 = 2.7 \times 10^{-6} ~;~g^2_2 = 3.73 \times 10^{-5}$, the time scale for decay of Dark-like mode ($\tilphi_1$) is $\simeq 14 $ times longer than that of the Higgs-like mode ($\tilphi_2$). For these parameters we find $|A_\phi|^2 = 4.88 \times 10^{-3}~;~|B_\phi|^2 = 0.871$, namely the probability for the Dark-like mode ($|A_\phi|^2$) is much smaller than that for the Higgs-like mode ($|B_\phi|^2$) .

For comparison,  figure (\ref{fig:fiproba12}) for $\mathcal{P}_{\phi \rightarrow \phi}$ also displays  Higgs decay without kinetic mixing  corresponding to $\varepsilon_R =0$. Fig. (\ref{fig:filogpro}) shows $\ln\Big(\mathcal{P}_{\phi \rightarrow \phi}(t)\Big) $ vs. $\tau_k$ to display more clearly the separation of scales in the almost degenerate case.   Note the scale on the horizontal axis, the oscillation period corresponds to $ \delta \tau_k \simeq  2\pi/\eta \simeq 10^3$, with $\tau_k $ given by eqn. (\ref{tau}) this oscillation time scale is $ \gtrsim 10^3$ \emph{longer} than the Higgs oscillation time scale $\simeq 1/\sqrt{\overline{M^2}}$. On this scale the natural Higgs oscillation occurs on a scale $\tau_k \simeq 1$ and the Higgs lifetime (without mixing) is $\tau_k \simeq 10^4$.  This is precisely the nature of the \emph{slow} time evolution captured by the effective Wigner-Weisskopf Hamiltonian with matrix elements (\ref{H11s}-\ref{H1221s}).

\vspace{2mm}

\subsubsection{$ \varepsilon_R \ll |\delta| \ll 1$:}\label{subsec:case13}  This nearly degenerate case   describes a broader region of parameters with $\varepsilon_R \ll \eta \ll 1$ allowing us to implement useful approximations and provide an analytic treatment.       As an example of this case, we note that for $M_\chi  =  130~,~135\,\mathrm{GeV}$ and $ \varepsilon_R = 10^{-3}$ one obtains $\eta \simeq 0.039~,~ 0.077$ respectively.  We find
\bea && y_1 \simeq F_1 \simeq \frac{\varepsilon_R}{2\eta} \ll 1 ~~;~~ y_2 \simeq F_2 \simeq 1 + \mathcal{O}(\varepsilon^2_R/\eta) \nonumber \\ && h_1 \simeq  1 + \mathcal{O}(\varepsilon^2_R/\eta) ~~;~~ h_2 \simeq -\frac{\varepsilon_R}{2\eta} \ll 1 \,, \label{intlimit}\eea thus we refer to $\tilphi_1$ as the Dark-like ($\chi$-like) mode and $\tilphi_2$ as the Higgs- like ($\phi$-like) mode,  with
\bea g^2_1 & \simeq & \frac{3\,\widetilde{Y}_R^2}{8\pi}\, \Big(\frac{\varepsilon_R}{2\eta}\Big)^2  \label{g1deg} \\
g^2_2 & \simeq &  \frac{3\,\widetilde{Y}_R^2}{8\pi} \,.\label{g2deg} \eea The result (\ref{g1deg}) is important: just on the basis of mixing with a parameter $\varepsilon_R$ one would expect that the decay rate for the $\chi$-like mode would be  suppressed by a factor $\varepsilon^2_R$ with respect to that of the $\phi$-like mode, however,  when the Higgs and Dark scalar are  nearly degenerate and $\varepsilon_R\ll 1$ there is an   \emph{enhancement} of the $\chi$-like rate by a factor $1/\eta^2 \gg 1 $.

In this  limit  there is a wide separation of the time scales of decay of the mass eigenstates because $g^2_1 \simeq g^2_2 \times (\varepsilon_R/2\eta)^2 \ll g^2_2$.

\begin{figure}[ht!]
\begin{center}
\includegraphics[height=4in,width=4.5in,keepaspectratio=true]{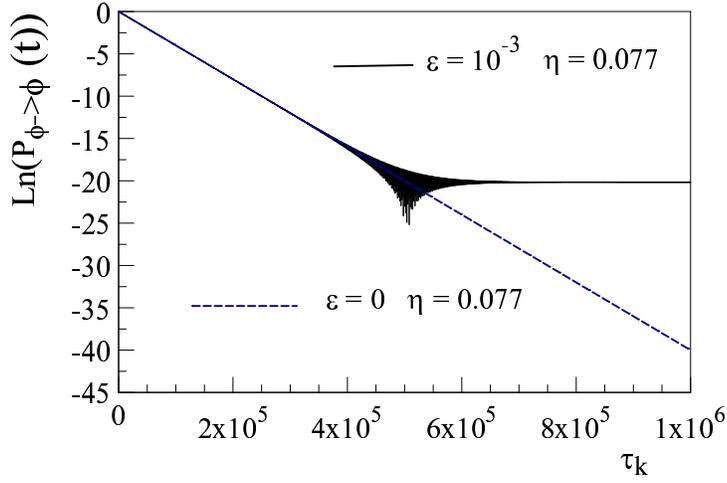}
\caption{ $\ln\Big(\mathcal{P}_{\phi \rightarrow \phi}(t)\Big) $, the solid line corresponds to $\varepsilon_R = 10^{-3}~,~\eta=0.077$, corresponding $M_\chi = 135\,\mathrm{GeV}$, $g^1_1 \simeq 2\times 10^{-9}, g^2_2 \simeq 4\times 10^{-5}$, the dashed line shows Higss decay without mixing ($\varepsilon_R=0$). At early times the Higgs-like mode dominates, at intermediate times the interference between Higgs and Dark-like modes dominates and the Dark-like mode dominates at late times (see text).  }
\label{fig:filogpro077}
\end{center}
\end{figure}

 Fig(\ref{fig:filogpro077})   shows $\ln\Big(\mathcal{P}_{\phi \rightarrow \phi}(t)\Big) $ vs. $\tau_k$  for $\varepsilon_R = 10^{-3}$   with $\eta = 0.077$ corresponding to $M_{\chi} = 135 \,\mathrm{GeV}$. For this case $g^2_1 \simeq 2\times 10^{-9}~;~g^2_2 \simeq 4 \times 10^{-5}$.   Larger values of $\eta$ lead to a wider separation of the time scales but also to faster oscillations which average out the oscillatory component. In the case of Fig(\ref{fig:filogpro077}) with $\varepsilon_R/\eta = 0.013 \ll 1$ and  $|A_\phi|^2 = 1.8\times 10^{-9} << |B_\phi|^2 \simeq 1$ the main features of the persistence probability (\ref{probas}) with $\alpha = \phi$ are much easier to understand: at early time the Higgs-like mode represented by  the direct term proportional to $|B_\phi|^2$ dominates  until a time   $\tau^*_k \simeq   \ln \big(\big|B_\phi / A_\phi\big|^2\big)/g^2_2$ at which this term becomes of the same order as the oscillatory term.  For $\tau_k > \tau^*_k$  the oscillatory term dominates, this is the interference between the Higgs and Dark-like modes, but this oscillatory contribution decays with $e^{-g^2_2 \tau_k/2}$ since $g^2_2 \gg g^2_1$, finally for $\tau_k \gg 2/g^2_2$ the Dark-like mode, represented by the direct term $|A_\phi|^2 \,e^{-g^2_1 \tau_k}$ dominates, this term is nearly constant up to $\tau_k \approx 1/g^2_1 \simeq   10^{10}$. In the units displayed in fig. (\ref{fig:filogpro077}) the Higgs lifetime is $\tau_k \simeq 2 \times 10^4$.

 Consider an intermediate time scale such that
\be 1/g^2_2 \ll  \tau_k \ll  1/g^2_1 \,, \label{inttau}\ee during this time scale $\mathcal{P}_{\phi \rightarrow \phi} \approx |A_\phi|^2 \simeq (\varepsilon/2\eta)^4 $ is nearly a constant and begins to decay on the much longer time scale $1/g^2_1$. This delay translates into a spatial pattern as a displaced decay vertex. This is similar to the phenomenon of regeneration in the neutral Kaon system which features a short and a long lived mass eigenstate\cite{commins}.

\subsection{Case II: $\eta \ll g^2_1,g^2_2$. } This case corresponds to
\be \frac{3\widetilde{Y}_R^2}{8\pi} \simeq 4 \times 10^{-5} \gg \eta \,,\label{yuklargeta}\ee this inequality also implies that both $\varepsilon_R, |\delta| \ll 10^{-5}$. Unless there is a fine tuning that makes $M_\chi \neq  M_\phi$  but keeping $|M_\chi-M_{\phi R}|/(M_\chi+M_{\phi R})\simeq 10^{-5}$, we assume that this parameter range corresponds to the degenerate case $\delta =0$, namely $\eta = \varepsilon_R$, again allowing useful approximations and an analytic treatment. We note that with $\eta =\varepsilon_R \neq 0$ although $M_\chi = M_{\phi R}$, the mass eigenstates are split by the kinetic mixing. With $\delta =0$ and $\eta= \varepsilon_R \ll 1$ we find  to leading order  that $y_1=y_2 = F_1 = F_2 =1/\sqrt{2}$, and
\be g^2_1 = g^2_2 = g_1g_2 \equiv g^2 =  \frac{3\widetilde{Y}_R^2}{16\pi}\,. \label{gis}\ee We recognize in this degenerate case that $\overline{M}^2 = M^2_{\phi R}  ~;~ \overline{E}_k = E_k = \sqrt{k^2+M^2_{\phi R}}$ and
\be g^2 \frac{M^2}{2E_k} = \frac{\Gamma_\phi(k)}{4}  \label{higsdeca}\ee where $\Gamma_\phi(k)$ is the Higgs decay rate into $\overline{f}f$  in the laboratory frame. It is convenient to introduce
\be \kappa = \frac{\varepsilon_R}{g^2} \ll 1 \,,\label{lam}\ee in terms of which we find to leading order
\be \lambda_+ = -i \,\frac{\Gamma_\phi(k)}{2} \,\Big(1-\frac{\kappa^2}{4} \Big)  ~~;~~ \lambda_- = -i \, \frac{\kappa^2 \Gamma_\phi(k)}{8} \,,  \label{lams}\ee along with
\be \mathds{C} = \frac{i\kappa}{\sqrt{1-\kappa^2}}~~;~~ \mathds{S} = \frac{1}{\sqrt{1-\kappa^2}}\,, \label{CS}\ee leading to
\be A_\phi \simeq 1 + \frac{\kappa^2}{4} ~~;~~ B_\phi \simeq - \frac{\kappa^2}{4} \,. \label{ABfi}\ee Therefore, $\lambda_+$ corresponds to the Higgs-like mode and $\lambda_-$ to the Dark-like mode.

  To leading order in $\kappa$ the real part of the eigenvalues $\lambda_\pm$ vanish, in terms of the poles in the propagator, this means that the real part of the complex energies is the \emph{same} for both, in this case the resonances are spectrally unresolved because the difference in the real parts is much smaller than the widths. This situation is similar to that in the $K0-\overline{K}0$ system where the real part of the difference in complex energies is of the same order as  the sum of the widths\cite{commins}.

\vspace{2mm}

\textbf{Space-time evolution: displaced vertices} Consider that in a collision a Higgs particle is created at an initial time $t=0$, this state will evolve in space and time as a coherent superposition of the (unstable) mass eigenstates. In order to study the space-time evolution let us restore the fast evolving phase $e^{-i \overline{E}_k t}$ in eqn. (\ref{fiti}) so the Higgs amplitude (\ref{fiti}) in momentum space is given by
\be   \tilphi(k,t)    =  \tilphi(k,0)\,  \Big[ A_{\phi}\,e^{-iE_+(k)\,t}\,e^{-\frac{\Gamma_+(k)}{2}\,t}+ B_{\phi}\,e^{-iE_-(k) \,t}\,e^{-\frac{\Gamma_-(k)}{2}\,t} \Big]  \,,\label{higgsamp} \ee  where
\be   \overline{E}_k + \lambda_\pm  \equiv  E_{\pm}(k) -i \frac{\Gamma_{\pm} (k)}{2} \,.  \label{Epm}\ee
  The space-time evolution is best understood in terms of a wave packet description\cite{sachs}. Consider that the initial amplitude describes a Gaussian wave packet narrowly localized in momentum space around a wavevector $\vec{k}_0$, namely
\be \phi(\vec{k},0) = \mathcal{N}\,e^{-\frac{(\vec{k}-\vec{k}_0)^2}{2\,\sigma^2_P}}  \label{gausswp}\ee where $\sigma_P \ll |k_0|$ is the width in momentum space and $\mathcal{N}$ a normalization factor. For a narrow wavepacket the Fourier transform in $k$ is performed by expanding the various quantities around $\vec{k}_0$ up to second order and carrying out the Gaussian integrals. We find
\be \phi(\vec{x},t) \simeq N  e^{i\,\vec{k}_0\cdot \vec{x}}~\Big[ A_\phi \, e^{-iE_+(k_0) t} \,e^{-\frac{\Gamma_+(k_0)}{2}\, t}\, e^{- \frac{(\vec{x}-\vec{V}_+(k_0) t)^2}{2\,\sigma^2_X} } + B_\phi \, e^{-iE_- (k_0)t} \,e^{-\frac{\Gamma_-(k_0)}{2}\, t}\, e^{-  \frac{(\vec{x}-\vec{V}_-(k_0) t)^2}{2\,\sigma^2_X}} \Big] \label{WPx} \ee where $\vec{V}_{\pm}(k_0)$ are the group velocities given by
\be \vec{V}_{\pm} = \frac{\vec{k}_0}{E_{\pm}(k_0)}\,, \label{groupvel}\ee $N$ is a normalization factor, $\sigma_X = 1/\sigma_P$ is the localization length in coordinate space. To obtain (\ref{WPx})  we have neglected the time dependence on $\sigma_X$ from dispersion and spreading, as well as subleading terms proportional to  $\widetilde{Y}_R^2$. This space-time amplitude describes \emph{two } localized wavepackets moving with slightly different  group velocities, the amplitude of each decays with the corresponding lifetimes $\Gamma_\pm(k_0)$.
The space-time persistence probability is given by
\be \mathcal{P}_{\phi \rightarrow \phi}(\vec{x},t) = |\phi(\vec{x},t)|^2 = |\phi_+(\vec{x},t)|^2 + |\phi_-(\vec{x},t)|^2 + I(\vec{x},t) \label{pfifixt}\ee where in obvious notation $|\phi_{\pm}(\vec{x},t)|^2 $ are the direct terms and  the interference contribution is given by
\bea && I(\vec{x},t)    =    N^2\,e^{- \frac{(\vec{x}-\vec{V}_+(k_0) t)^2}{2\,\sigma^2_X} }\,e^{-  \frac{(\vec{x}-\vec{V}_-(k_0) t)^2}{2\,\sigma^2_X}}\,e^{-\frac{\Gamma_+(k_0)+\Gamma_+(k_0)}{2} t} \times \nonumber \\ &   &  \Bigg[ \big[ A_{R \phi} B_{R \phi} + A_{I \phi} B_{I \phi}] \cos\Big[(E_+-E_-) \,t\Big] + \big[ A_{I \phi} B_{R \phi} - A_{R \phi} B_{I \phi}] \sin\Big[ (E_+-E_-) \,t \Big]\Bigg] \,.\label{interference} \eea This interference term is a hallmark of coherence, which is suppressed by \emph{two} sources of decoherence: i) the decay with the average of the rates of the two mass eigenstates and ii) decoherence from the separation of wavepackets and suppression of their overlap. This latter effect arises from the fact that the wave packets evolve with slightly different group velocities   separating  as time evolves. The overlap vanishes when the distance between the centers   is larger than the width of the wave packets, this happens for $t > t_{coh}$ where
\be t_{coh} \simeq \frac{\sigma_X}{\Big|\vec{V}_-(k_0)- \vec{V}_+(k_0)\Big|}\,.  \label{tcohe}\ee
 This latter source of decoherence is similar to the decoherence of neutrino oscillations in the wavepacket formulation\cite{nuwp,beuthe,giunti}.

\vspace{2mm}

\textbf{Case I:  ($\eta \gg 10^{-5}$)}  In this case the results (\ref{lamplu},\ref{lammin}) imply that to leading order $E_+(k) = \sqrt{k^2+M^2_1}~~;~~E_-(k) = \sqrt{k^2+M^2_2}$, therefore the difference in group velocities becomes
\be \Big|\vec{V}_-(k_0)- \vec{V}_+(k_0)\Big| = \frac{ {k_0}}{\overline{E}(k_0)} \frac{\overline{M^2}}{  \overline{E}^2(k_0)}\, \eta \,.\label{veldif} \ee The time scale for decoherence by wavepacket separation is given by
\be   t_{coh} \simeq  \frac{\sigma_X\,{\overline{E}}^{\,3}(k_0)}{k_0\,\overline{M^2}\,\eta} \label{cohtimecase1}\ee  Therefore the interference term vanishes at a time scale $t_{dec}$ given by  the \emph{smallest}  between $t_{coh}$ and the inverse of the average decay rate of the two mass eigenstates. For $t \gg t_{dec}$  only the direct contribution from the mass eigenstate with the longest lifetime  survives in the persistence probability. For $\varepsilon_R \leq \eta$ this is the mass eigenstate corresponding to $\lambda_+$, namely for $t \gg t_{dec}$
\be \mathcal{P}_{\phi \rightarrow \phi}(\vec{x},t) \simeq  |\phi_+(\vec{x},t)|^2 = N^2 \, |A_\phi|^2 \,e^{-\frac{(\vec{x}-\vec{V}_+(k_0) t)^2}{\sigma^2_X}}\,e^{-\Gamma_+(k_0)\, t}\,. \label{dispprob} \ee

For $\Gamma_+ \ll \Gamma_-$ this probability describes a  decay vertex displaced from the origin by a distance
\be  \Delta x  \approx \frac{\big|V_+(k_0)\big|}{\Gamma_+(k_0)}\,.  \label{displace} \ee
 In the broad parameter range with $|\delta| \gg \varepsilon_R $ we can use the results (\ref{g1deg}) to find that
\be  \Delta x
\simeq \big|V_+(k_0)\big| \,\frac{\overline{\gamma}(k_0)}{\Gamma_\phi}\,\Big(\frac{2\, \eta}{\varepsilon_R}\Big)^2\,, \label{dispdist}\ee where $\Gamma_\phi$ is the Higgs decay rate (at rest)  in the $\overline{f}f$ channel and we used eqn. (\ref{intlimit}) in the limit $\varepsilon_R/\eta \ll 1$. This spatial separation suggests a potential telltale signature of kinetic mixing: two well separated decay vertices into the \emph{same} channels, the first corresponding to the decay of the Higgs-like mode $\tilphi_2$ and the second displaced by a distance given by (\ref{dispdist}) from the decay of the Dark-like mode $\tilphi_1$. Taking  $\Gamma_\phi = 4\,\mathrm{MeV}$ as  the total decay width of the Standard Model Higgs (see section (\ref{sec:discussion})), we  find
\be  \Delta x  \approx 2 \, \Bigg[\frac{\big|V_+(k_0)\big|  \overline{\gamma}(k_0)}{10} \Bigg]\,\Big[ \frac{|\delta|}{10^{-1}}\Big]^2\,   \Big[\frac{10^{-6}}{\varepsilon_R}\Big]^2\,\mathrm{cm}\,. \label{deltax} \ee

 In the limit  $\varepsilon_R/\eta \ll 1$, namely $\eta \simeq |\delta|$,  the relative probability for the Dark-like component to decay with a displaced vertex is given by
\be  \mathcal{R} \simeq \frac{|A_\phi|^2}{|B_\phi|^2} \simeq  \, \Big[\frac{10^{-1}}{\delta} \Big]^4\,\Big[ \frac{\varepsilon_R}{10^{-6}}\Big]^4 \, \times 10^{-21} \,. \label{BR}\ee Thus we see that a larger displacement of the vertex implies a much smaller probability, this trade-off between the decay length and the probability  suggests a challenging observational scenario for displaced vertices.
As an example consider $M_\chi = 135 \mathrm{GeV}~;~ V \overline{\gamma} \sim 10$ and $\varepsilon_R = 10^{-6}$ yielding $\eta \simeq |\delta| \simeq 0.077$,   $ \Delta x  \approx 1\,\mathrm{cm}$ and $\mathcal{R} \simeq 10^{-21}$.

\vspace{2mm}

\textbf{Case II: ($\eta \ll 10^{-5}$)} Considering the degenerate case $\delta =0$, the results  (\ref{lams}) and (\ref{ABfi})  yield to leading order
\be \mathcal{P}_{\phi \rightarrow \phi}(\vec{x},t)   = N^2 \,  \,e^{-\frac{(\vec{x}-\vec{V}(k_0) t)^2}{\sigma_X}}\,\Bigg[\big(1+\frac{\kappa^2}{4} \big)\,e^{-\frac{\Gamma_\phi(k_0)}{2}\, t} - \frac{\kappa^2}{4}\,e^{-\frac{\kappa^2 \Gamma_\phi(k_0)}{8}\, t} \Bigg]^2\,. \label{dispprobcase2} \ee The first  and second terms in the bracket  are identified as the Higgs-like and Dark-like modes respectively. In this case the interference term is not oscillatory because to leading order in $\kappa$ the eigenvalues $\lambda_\pm$ are purely imaginary. To this order the difference $E_+ - E_- \simeq 0$. This case is similar to $K_0 \overline{K}_0$ mixing where the real part of the difference in the eigenvalues is smaller than (or of the same order as) the difference in the imaginary parts and the interference term does not feature an oscillatory component\cite{commins,cppuzzle,cpviolation}.  For $\kappa \ll 1$ the Dark-like mode dominates for $t > -4 (\ln\kappa)/ \Gamma_\phi(k_0) $  leading to a late decay with a vertex displaced by
\be \Delta x \simeq 8 \,\Bigg[\frac{\big|V_+(k_0)\big|  \overline{\gamma}(k_0)}{10} \Bigg]\, \Bigg[\frac{10^{-10}}{\varepsilon_R}  \Bigg]^2 \, \mathrm{cm}\,, \label{delx} \ee with a ratio of probabilities (for $\varepsilon_R \ll 2\times 10^{-5}$)
\be \mathcal{R} \simeq  \Bigg[\frac{\varepsilon_R}{4\times 10^{-5}} \Bigg]^4 \,.\label{smalBR}\ee

Again, as in the previous case there is a trade-off between a larger displacement and a smaller probability.

\section{Radiative corrections from radion-(SM) coupling.}\label{sec:radionsm}
As discussed in section (\ref{sec:diag}) within the radion model, the radion features couplings to the standard model degrees of freedom, its coupling to fermions and gauge vector bosons is given by eqn. (\ref{LInt}).
We now extend the discussion of the previous sections by including the radiative corrections from this coupling. Following the steps in section (\ref{sec:effac}) leading to the effective action we now find up to one fermionic and vector boson loop in momentum space,
\be \delta S_{eff}[\phi] = -\frac{1}{2}\int \frac{d^4p}{(2\pi)^4} \Big\{\tilphi(-p)\,\tilsigma_{\phi\phi}(p)\, \tilphi(p) +   \tilchi(-p)\,\tilsigma_{\chi\chi}(p)\, \tilchi(p)  + 2\,\tilphi(-p)\,\tilsigma_{\phi\chi}(p)\, \tilchi(p)   \Big\}   \,, \label{delrsm} \ee where
\be \tilsigma_{\phi\phi}(p) = \tilsigma(p)~~;~~\tilsigma_{\chi\chi}(p) = \gamma^2 \,\tilsigma(p)~~;~~
\tilsigma_{\phi\chi}(p) = \gamma  \,\tilsigma(p)\,, \label{sigmasrsm}\ee and now $\tilsigma(p)$ is the one loop self energy including  fermions and vector bosons. Anticipating the necessity for an off-diagonal mass term for renormalization, we add a counterterm $-\chi \, m^2_{\phi\chi} \, \phi$ to the bare action, leading to the the total one loop effective action   given by
\bea S_{eff}[\phi,\chi]  & = &  \int \frac{d^4 p}{(2\pi)^4} \Bigg\{ \frac{1}{2} \,\tilphi(-p)\,\big[p^2-M^2_\phi-\tilsigma_{\phi\phi}(p) \big]\tilphi(p)+ \frac{1}{2} \,\tilchi(-p)\,\big[p^2-M^2_\chi-\tilsigma_{\chi\chi}(p) \big]\tilchi(p) \nonumber \\ & & -  \, \tilchi(-p)\Big[\varepsilon \, p^2 +m^2_{\phi\chi}+ \tilsigma_{\phi\chi}(p) \Big]  \,\tilphi(p) \Bigg\}\,. \label{totSeffrasm}\eea As discussed in section (\ref{sec:effac}) the real part of the one loop self energy requires two substractions, for the diagonal self-energies we subtract on the (renormalized) mass shells (before mixing), namely
\bea   \mathrm{Re}\Big[\tilsigma_{\phi\phi}(p^2)\Big] & = &  \mathrm{Re}\Big[\tilsigma_{\phi\phi}(M^2_{\phi R})\Big] + (p^2-M^2_{\phi R})\,\mathrm{Re}\Big[\tilsigma'_{\phi\phi}(M^2_{\phi R})\Big] + \mathrm{Re}\Big[\tilsigma^f_{\phi\phi}(p^2)\Big] \label{subphi}  \\  \mathrm{Re}\Big[\tilsigma_{\chi\chi}(p^2)\Big] & = &  \mathrm{Re}\Big[\tilsigma_{\chi\chi}(M^2_{\chi R})\Big] + (p^2-M^2_{\chi R})\,\mathrm{Re}\Big[\tilsigma'_{\chi\chi}(M^2_{\chi R})\Big] + \mathrm{Re}\Big[\tilsigma^f_{\chi\chi}(p^2)\Big] \,,\label{subchi}  \eea whereas for the off-diagonal term we subtract at $p^2=\overline{M^2}$,
\be    \mathrm{Re}\Big[\tilsigma_{\phi\chi}(p^2)\Big]   =    \mathrm{Re}\Big[\tilsigma_{\phi\chi}(\overline{M^2})\Big] +  (p^2 - \overline{M^2}) \,\mathrm{Re}\Big[\tilsigma'_{\phi\phi}(\overline{M^2})\Big] + \mathrm{Re}\Big[\tilsigma^f_{\phi\chi}(p^2)\Big] \,,  \label{subphichi} \ee  so that $\mathrm{Re}\Big[\tilsigma^f_{\phi\chi}(\overline{M^2})\Big] = 0 $ and is finite for all $p^2$.  The self energies $\Sigma^f = \mathrm{Re}\Sigma^f + i \mathrm{Im}\Sigma$ are finite. The renormalization conditions for the diagonal components follow eqns. (\ref{massren}-\ref{zphi}) for $\phi,\chi$ respectively, whereas for the off-diagonal component they become
\bea && m^2_{\phi\chi}+ \mathrm{Re}\Big[\tilsigma_{\phi\chi}(\overline{M^2})\Big]- \overline{M^2}  \,\mathrm{Re}\Big[\tilsigma^{'}_{\phi\chi}(\overline{M^2})  \Big]   =   {\tilde{m}_{\phi\chi}}^2 \nonumber \\ && \varepsilon + \mathrm{Re}\Big[\tilsigma'_{\phi\chi}(\overline{M^2}) \Big]   =   \tilde{\varepsilon}\,.   \label{offrens}\eea The first equation in (\ref{offrens}) justifies the addition of the off-diagonal mass counterterm to absorb a divergence of the one loop self energy,  and $\tilde{\varepsilon}$ is an intermediate renormalization. The interesting aspect of this analysis is that including radiative corrections suggests that the most general tree level action must include an off-diagonal mass term to renormalize the divergences in the self-energies. Since our focus is in understanding the consequences of kinetic mixing, we set the intermediate parameter ${\tilde{m}_{\phi\chi}}^2 =0$ by adjusting the counterterm to cancel this self energy subtraction. The possibility of a non-vanishing   off-diagonal (renormalized) mass term while interesting is beyond the scope of this article and is postponed to further study. Finally the renormalization of the kinetic mixing is achieved by
\be \varepsilon_R = \sqrt{Z_\phi\,Z_\chi}\,\,\tilde{\varepsilon} \,. \label{renmixrsm}\ee Renormalizing the couplings by multiplicative renormalization with the wave function renormalization constants also suggests a renormalization of the parameter $\gamma$ that determines the couplings of the radion to the (SM) degrees of freedom, which is neglected to the order that we study. The final form of the effective action  after renormalization of couplings is now given by
\bea && S_{eff}[\phi,\chi]   =   \int \frac{d^4 p}{(2\pi)^4} \Bigg\{ \frac{1}{2} \,\tilphi_R(-p)\,\big[p^2-M^2_{\phi R}\big]\tilphi_R(p)+ \frac{1}{2} \,\tilchi_R(-p)\,\big[p^2-M^2_{\chi R} \big]\tilchi_R(p) -   \varepsilon_R \,\, \tilchi_R(-p)\,p^2 \,\tilphi_R(p)   \nonumber \\   & & -    \frac{1}{2}   \,  \tilphi_R(-p)\,\,\tilsigma^f_{\phi \phi}(p^2) \,\,\tilphi_R(p)-\frac{1}{2} \,  \tilchi_R(-p)\,\,\tilsigma^f_{\chi \chi}(p^2) \,\,\tilchi_R(p) -\tilchi_R(-p)\,\,\tilsigma^f_{\phi \chi}(p^2) \,\,\tilphi_R(p) \Bigg\} \,.
\label{totSeffrenrsm}\eea We now proceed in exactly the same way as in section (\ref{sec:effham}) by diagonalizing the first line (renormalized tree-level) and writing the second line in terms of
\be \tilphi_R(p) = y_1 \tilphi_1(p) + y_2 \tilphi_2 (p)~~;~~ \tilchi_R(p) = h_1 \tilphi_1(p) + h_2 \tilchi_2 (p) \label{relanew} \ee where $y_{1,2},h_{1,2}$ are given by the same expressions as in section (\ref{sec:diag}) in terms of the renormalized parameters.  We note that
\be \tilsigma^f_{\phi \phi}(p^2) \propto Y^2, \alpha^2_w ~~;~~ \tilsigma^f_{\chi \phi}(p^2) \propto \gamma Y^2, \gamma \alpha^2_w ~~;~~ \tilsigma^f_{\chi \chi}(p^2) \propto \gamma^2 Y^2, \gamma^2 \alpha^2_w \,,\label{magssig}\ee therefore to leading order in $\gamma \ll 1$ we neglect the  term in $\tilde{\chi}_R \,\tilsigma^f_{\chi \chi}\,\tilde{\chi}_R$ in the effective action. We obtain the effective Hamiltonian following the same steps as in section (\ref{sec:effham}). The renormalization prescriptions described above entail  that the real part of the diagonal self energies are subleading as in section (\ref{sec:effham}) and that of the off-diagonal term  vanishes when evaluated at the scale $\overline{M^2}$. Using (see eqns (\ref{y1y2},\ref{h1h2})) which to leading order in $\varepsilon_R$ become
\be h_1 = y_2 ~~;~~ h_2 = -y_1 \label{relasrsm}\ee and that the imaginary part of the one-loop self-energy on the (near) mass shell scale $\overline{M^2}$ only receives a leading contribution from the fermion-antifermion loop, it follows that
\be \mathrm{Im}\Big[\tilsigma^f_{\phi\phi}(\overline{M^2}) \Big] = -\frac{3\,\widetilde{Y}_R^2}{8\pi}\, \overline{M^2} ~~;~~\mathrm{Im}\Big[\tilsigma^f_{\phi\chi}(\overline{M^2}) \Big] = \gamma \, \mathrm{Im}\Big[\tilsigma^f_{\phi\phi}(\overline{M^2}) \Big]\,. \label{imagfichi}\ee To leading order in $\gamma$ we find that the matrix elements of the effective Wigner-Weisskopf Hamiltonian (\ref{H11s}),(\ref{H22s}),(\ref{H1221s})  now become
 \bea H_{11} & = &  \frac{\overline{M^2}}{2 \overline{E}_k}\,\big[ \eta - i\,G^2_1\big]   \label{H11srsm} \\ H_{22} & = &  \frac{\overline{M^2}}{2 \overline{E}_k}\,\big[ -\eta-i\,G^2_2  \big] \label{H22srsm} \\
 H_{12} & = & H_{21} =  \,\frac{\overline{M^2}}{2 \overline{E}_k}\big[-i\, G_1\,G_2\,\big] \,. \label{Hwwrsm} \eea where to leading order in $\gamma$
 \be G_1 = g_1 + \frac{\gamma}{2} \, g_2 ~~;~~ G_2 = g_2-\frac{\gamma}{2} \,g_1 \,.\label{newww}\ee  We are now in position to assess the contribution from the radion couplings to the standard model degrees of freedom in the cases analyzed in section (\ref{sec:timevol}) in the parameter range $\varepsilon_R \simeq \gamma \ll 1 $ as suggested by the   analysis of LHC data in ref.\cite{chak}. For $\varepsilon_R \simeq \eta \ll 1$ it follows that $g_1 \simeq g_2$ and the terms proportional to $\gamma \ll 1$ can be safely neglected. In the case $\varepsilon_R \ll \eta \ll 1$ the results from eqn. (\ref{intlimit}) along with $g_1/g_2 = \varepsilon_R/2\eta$  yield $G_2 \simeq g_2$ and $G_1 = g_1 \,(1+\eta \, \gamma/2\varepsilon_R)$ which for $\gamma \simeq \varepsilon_R  $ and $\eta \ll 1$ entails that to leading order $G_1=g_1$, thus in this case the contribution from $\gamma$ can also be neglected. Finally in the case $\eta \ll g^2_1\,,\,g^2_2$ with $y_1 \simeq y_2 $ the contribution from $\gamma$ can also be neglected. Therefore we conclude that in all the cases of interest analyzed in the previous section, and in the parameter range $\varepsilon_R \simeq \gamma$ the contributions from the couplings between the radion and the standard model degrees of freedom can be safely neglected.

\section{Discussion:}\label{sec:discussion}

\textbf{i:) Generalization.} We have obtained the effective action and effective Hamiltonian by considering solely a Yukawa coupling between the Higgs and fermionic degrees of freedom of the Standard Model to highlight the main conceptual steps. In the previous section we also included the coupling of the radion to (SM) degrees of freedom.  However, we can simply generalize the one loop effective action by integrating out all the degrees of freedom that couple to the Higgs and contribute to the    Higgs self-energy. All of these will contribute to mass and wave function renormalization, however, only those corresponding to intermediate states with multiparticle thresholds \emph{below} the position of the renormalized mass of the Higgs will contribute to the imaginary part of the effective Hamiltonian. We can simply include these in the one loop effective Hamiltonian  by replacing
\be   \frac{3\,\widetilde{Y}_R^2}{8\pi} \rightarrow \frac{\Gamma_H}{M_{\phi R}} \label{replace1lup}\ee where $\Gamma_H = 4 \,\mathrm{MeV}$ is, now,  the \emph{total} Higgs decay width. At the level of the propagator, it is tantamount to replacing the one-loop self-energy with the full self-energy, in principle to all orders in Standard Model couplings.

\vspace{2mm}

\textbf{ii:) Effective action vs.   Wigner-Weisskopf method: }
In the Wigner-Weisskopf approach to neutral meson mixing, one obtains a Schroedinger like equation for the amplitudes in the interaction picture by taking transition matrix elements of  the interaction Hamiltonian (in the interaction picture)\cite{lee}-\cite{boya}.  Truncating the hierarchy of equations to the states that are connected to the initial state by the interaction Hamiltonian at a given order (typically second order)  yields an effective Schroedinger equation for the amplitudes with a non-Hermitian Hamiltonian (for a detailed discussion see refs.\cite{cpviolation,cppuzzle,boya}).  The interaction via kinetic mixing does not lend itself \emph{directly} to such treatment. One can perform a field redefinition and diagonalization of the resulting mass matrix as described in section (\ref{sec:diag}) above, in which case the Standard Model vertices in terms of the mass eigenstates can be taken as the interaction Hamiltonian. However, while it is straightforward to extract mass renormalization in this description, it is less clear how to include wave-function renormalization, which is a consequence of the  momentum dependent ultraviolet divergence of  the fermionic self-energy.  Furthermore, as discussed in section (\ref{sec:effac}) renormalization is more cumbersome  in the mass basis since there are several renormalization conditions. The method described in sections (\ref{sec:effac}, \ref{sec:effham})  in terms of the equations of motion from the effective action  unambiguously and straightforwardly leads to the  Schroedinger-like equation for the renormalized amplitudes in terms of a non-Hermitian Hamiltonian with the fully renormalized masses, field amplitudes and decay widths.

\vspace{2mm}

\textbf{iii:) Kinetic vs. mass mixing:} After the transformation (\ref{alfabeta}) and the field redefinition (\ref{rescale}) the Lagrangian at the quadratic level is identical to two massive fields coupled bilinearly with a mass mixing term (\ref{ABlag}). Diagonalizing the mass matrix yields the mass eigenstates. Although at first sight one would infer that kinetic mixing is equivalent to mixing with an off-diagonal mass matrix, in fact they are different. The main difference is that in the case of kinetic mixing, the mass eigenstates are \emph{not} unitarily related to the  original fields  because of the rescaling (\ref{rescale}), whereas in mass mixing, the fields that create the mass eigenstates are unitarily related to the original fields.

\vspace{2mm}

\textbf{iv:) Bounds on radion-Higgs mixing:} Earlier analysis of LHC data\cite{desai,sandes} revealed that the constraints for radion-Higgs mixing with near degeneracy loosen substantially for very small mixing and $v/\Lambda_\chi \ll 1$. The most recent constraints\cite{chak} confirm and tighten the earlier bounds, showing an allowed region of parameter space for small mixing and $v/\Lambda_\chi\ll 1$. In particular the most interesting region for displaced vertices discussed in the previous section corresponds to $\varepsilon/\eta \ll 1$, which when combined with $v/\Lambda_\chi \ll 1$ corresponds to the ``conformal point'' investigated in detail in ref.\cite{chak}. For this region of parameters, ref.\cite{chak} finds a wedge of allowed region for radion-Higgs mixing which widens as $\Lambda_\chi$ increases. This is precisely the region of parameters in which our theoretical analysis focused, thus lending support to the possibility that the dynamical effects found in the previous sections could be experimentally relevant.

\textbf{v:) Displaced vertices at the LHC:} The Atlas detector at the LHC reported  results of a search program  for the decay of long-lived neutral particles with displaced vertices  at $\sqrt{s} = 7, 8 \,\mathrm{TeV}$\cite{atlas}. No significant excess of events over background were found. The benchmark scenario used for the analysis  which is more relevant to the discussion here, is that of ref.\cite{zurek}. In this scenario the Higgs mixes with a scalar (but not kinetically) which in turn decays into a pair of new hadronic states, each one finally decaying into $\overline{b}b$ pairs via a new gauge boson $Z'$ with a displaced vertex.

The scenario studied in the previous sections  is very different in that the initial Higgs is a coherent superposition of a Higgs-like and a Dark-like state, with nearly the same masses but very different lifetimes. This model is defined by only two parameters: the strength of the kinetic coupling $\varepsilon_R$ and the mass of the Dark scalar (radion) (encoded in the degeneracy parameter $\delta$). The benchmark model used for analysis of the LHC data\cite{zurek} features a much richer dark sector with various new hadronic resonances and gauge bosons with various new couplings and masses, therefore introducing more parameters, interactions and model dependence than in the model that we study here.

Although experimentally both scenarios may yield the same final state ($\overline{b}b \rightarrow \mathrm{jets}$), physically the processes are very different, hence using analysis based on   the benchmark   model of ref.\cite{zurek} may not be suitable for the mixing/oscillation case studied here. The results for the displaced vertex (\ref{deltax}) or (\ref{delx})  and relative probability (\ref{BR})  or (\ref{smalBR}) for cases I, II respectively,  suggest that detection of displaced vertices will be very challenging.

\section{Conclusions and further questions:}
Motivated by higher dimensional extensions beyond the Standard Model, in this work we studied the \emph{dynamical} aspects of   kinetic mixing between the Higgs particle and a nearly degenerate $SU(3)_c\times SU(2)\times U_Y(1)$ singlet Dark scalar field. This is an alternative portal to degrees of freedom beyond the Standard Model that may possibly be suitable  Dark matter candidates. We focused on very small kinetic mixing (motivated by weakly interacting Dark matter) and a Dark scalar field \emph{nearly degenerate} with the Higgs field, a region of parameter space that has not yet been excluded and is weakly constrained by   LHC data\cite{chak,desai,sandes}.

 A further motivation is the realization that kinetic mixing offers a fundamentally different scenario of  mixing phenomena which, by itself,    merits a deeper study  as a complement to more phenomenological studies.

 One of the main results of this work is the implementation of  the renormalization program directly from  the effective action. This is  a new method that yields straightforwardly and unambiguously   fully renormalized Schroedinger-like equations of motion for the amplitudes in terms of a  non hermitian  effective Hamiltonian similar to the treatment of neutral meson mixing.

 Although we focused on a Yukawa coupling of the Higgs field to fermionic degrees of freedom of the Standard Model  we argue that the method  is more general.

Kinetic mixing of the Higgs with a nearly degenerate Dark scalar implies that the Higgs field is a coherent superposition of the mass eigenstates leading to oscillations and a common decay channel, again similarly to neutral meson mixing.
Small kinetic coupling and nearly degenerate Dark and Higgs scalars imply that while the mass differences of the mass eigenstates are small, their lifetimes are very different. These features have important implications: in a collision experiment producing a Higgs particle, the initial state evolves as a coherent superposition of the mass eigenstates, leading to interference  as a consequence of mixing an decay of the different components with a wide difference of the decay rates. The large difference in the lifetimes of the mass eigenstates implies a delayed decay of the Dark-like component with a displaced vertex \emph{into the same channels} as those available for Higgs decay.

There are two relevant dimensionless parameters: $\varepsilon_R$ is the renormalized kinetic mixing parameter (\ref{kinmixssb}) and $\delta$ given by (\ref{dege}) is a degeneracy parameter from which the combination $\eta= \sqrt{\varepsilon^2_R+\delta^2}$ plays a fundamental role in the description of the dynamics. We find two distinct regions of parameters, i) for  $\eta \gg \Gamma_H/M_H$, where $\Gamma_H, M_H$ are the width and mass of the Higgs, interference between the Dark-like and Higgs-like components give rise to oscillations in the ``persistence'' probability to find the Higgs-like mode, and for $\varepsilon_R \ll \eta \ll 1$ there is an enhancement of the effective kinetic coupling as a consequence of the near degeneracy. ii) for $\eta \ll  \Gamma_H/M_H$ the interference term does not feature oscillations as the oscillation frequency  is much smaller than the average width. In both cases we find a wide separation of lifetimes between the Higgs-like and Dark-like modes. A wave-packet treatment yields a space-time description of the delayed decay of the Dark-like mode with displaced vertices albeit with large vertex displacements correlated with very small probabilities presenting a challenging observational scenario.

The phenomena discussed in this study, namely interference effects and displaced vertices, must be input in the analysis of collider data to establish more firmly in forthcoming experiments the possibility of Dark scalar mixing kinetically with the Higgs sector in the region of parameter space studied here. While no excess signal over background has yet been found for displaced vertex events at the LHC, the search will continue with Run II and future linear colliders.

Kinetic mixing provides a novel scenario for mixing, oscillations and displaced vertices opening new theoretical and experimental possibilities.

\acknowledgements The authors  gratefully   acknowledge  support from NSF through grant PHY-1506912.

\end{document}